\documentclass[fleqn,usenatbib]{mnras}

\usepackage[T1]{fontenc}
\usepackage{ae,aecompl}

\usepackage{graphicx}	
\usepackage{amsmath}	
\usepackage{amssymb}	
\usepackage{multicol}
\usepackage{xspace}
\usepackage{xcolor}
\usepackage{booktabs}
\usepackage[normalem]{ulem}
\usepackage{newtxtext,newtxmath}

\def\sec#1{Sec.~\ref{sec:#1}}
\def\app#1{App.~\ref{sec:#1}}
\def\Fig#1{Fig.~\ref{fig:#1}}
\def\Table#1{Table~\ref{tab:#1}}
\def\Eq#1{Eq.~(\ref{eq:#1})}
\newcommand{\agn}{NIHAO (SF+BH)\xspace}
\newcommand{\noagn}{NIHAO (SF)\xspace}

\newcommand{\fdm}{$\mathcal{F_{DM}}$\xspace}
\newcommand{\rz}{\,$z=0$\xspace}

\title[The dark balance]{The dark balance: quantifying the inner halo response to active galactic nuclei feedback in galaxies}
\author[N. Arora et al.]
{Nikhil Arora,$^{1,2,3}$\thanks{E-mail: nikhil.arora@queensu.ca}
Stéphane Courteau,$^{3}$
Andrea V. Macci\`o,$^{1,2,4}$
Changhyun Cho,$^{1,2}$
\newauthor
Raj Patel$^{3}$ and
Connor Stone$^{5,6,7}$
\\
$^{1}$New York University Abu Dhabi, PO Box 129188, Abu Dhabi, United Arab Emirates\\
$^{2}$Center for Astrophysics and Space Science (CASS), New York University Abu Dhabi, Abu Dhabi, PO Box 129188, Abu Dhabi, UAE\\
$^{3}$Department of Physics, Engineering Physics \& Astronomy, Queen's University, Kingston, ON K7L 3N6, Canada\\
$^{4}$Max-Planck-Institut für Astronomie, Königstuhl 17, D-69117 Heidelberg, Germany\\
$^{5}$ Department of Physics, Universit{\'e} de Montr{\'e}al, Montr{\'e}al, Qu{\' e}bec, Canada\\
$^{6}$ Mila - Qu{\' e}bec Artificial Intelligence Institute, Montr{\'e}al, Qu{\' e}bec, Canada\\
$^{7}$ Ciela - Montr{\'e}al Institute for Astrophysical Data Analysis and Machine Learning, Montr{\'e}al, Qu{\' e}bec, Canada\\
}

\date{Accepted XXX. Received YYY; in original form ZZZ}

\pubyear{2023}

\begin{document}
\label{firstpage}
\pagerange{\pageref{firstpage}--\pageref{lastpage}}
\maketitle
 
\begin{abstract}
   This paper presents a study of the impact of supermassive black hole (SMBH) feedback on dark matter (DM) halos in numerical NIHAO simulations  of galaxies.
   In particular, the amount of DM displaced via active galactic nuclei (AGN) feedback and the physical scale over which AGN feedback affects the DM halo are quantified by comparing NIHAO simulations with and without AGN feedback.  
   NIHAO galaxies with $\log(M_*/M_{\rm \odot})\geq 10.0$ show a growing central DM suppression of 0.2 dex ($\sim$40\,per\,cent) from $z=1.5$ to the present relative to noAGN feedback simulations.
   The growth of the DM suppression is related to the mass evolution of the SMBH and the gas mass in the central regions.
   For the most massive NIHAO galaxies $\log(M_*/M_{\rm \odot}) > 10.5$, partially affected by numerical resolution, the central DM suppression peaks at $z=0.5$, after which halo contraction overpowers AGN feedback due a shortage of gas and, thus, SMBH growth.
   The spatial scale, or ``sphere of influence,'' over which AGN feedback affects the DM distribution decreases as a function of time for MW-mass galaxies (from $\sim$16\,kpc at $z=1.5$ to $\sim$7.8\,kpc at $z=0$) as a result of halo contraction due to stellar growth. 
   For the most massive NIHAO galaxies, the size of the sphere of influence remains constant ($\sim$16\,kpc) for $z>0.5$ owing to the balance between AGN feedback and halo contraction.
\end{abstract}



\section{Introduction}
\label{sec:Intro}

In a dark energy-cold DM dominated ($\Lambda \rm{CDM}$) hierarchical Universe, galaxy evolution is a complex interplay between gravitational and baryonic processes \citep{White1978, White1991}.
With most of the baryons in galaxies found in their central parts (at the minimum of the gravitational potential), the mass budget of all galaxies is dark matter (DM) dominated in the outskirts. 
Observations and simulations of galaxies have indeed shown that the fraction of DM $f_{\rm DM}(R)=M_{\rm DM}(R)/M_{\rm tot}(R)$ at large galactocentric radii approaches unity
\citep{Deason2012, Courteau2014, Remus2017, Harris2020}. 
However, at small galactocentric radii, the characterization of $f_{\rm DM}(R)$ is closely tied to the stellar mass concentration and various hydrodynamic processes \citep{CourteauRix1999, Dutton2011, Courteau2015} and remains under contention.

The fraction of baryon-to-DM in the inner parts of galaxies is strongly correlated to the stellar and total masses (by default), as well as environment and hydrodynamical (e.g., stellar and/or AGN feedback) processes. 
At the low mass end of the stellar mass function, feedback due to stellar winds and supernovae should be dominant \citep{DekelSilk1986, Katz1996, Stinson2006, Hopkins2012} while supermassive black holes (SMBH) are more effective in high mass galaxies \citep{Silk2012, Fabian2012} for altering galaxy formation and evolution.
Independently of stellar mass and redshift, these feedback mechanisms (both stellar and AGN) have been known to generate significant gas outflows \citep{Silk1998, Hopkins2012, Nelson2019} and to alter numerous galaxy properties such as star formation rates \citep{Baldry2006, Wilman2010, Fossati2015, Croton2016, Arora2019}, gas content \citep{Shangguan2018}, stellar and gas kinematics \citep{Frosst2022, Waterval2022}, and chemical enrichment \citep{Planelles2014, Huang2022}.

The feedback mechanisms alluded to above can also affect the local gravitational potential (or the overall mass distribution) through alteration of the baryonic mass distribution.
The radial distribution of the DM in galaxies is indeed expected to react to gravitational potential fluctuations, potentially leading to halo expansion and variations of the central DM density \citep{Dutton2007,Oh2011,Trujillo-Gomez2011,Pontzen2012,Teyssier2013,Dutton2016}.
For low-mass galaxies, supernovae driven outflows can alter the cuspy nature of DM halos leading to core creation~\citep{Governato2010, Teyssier2013, DiCintio2014, Chan2015, Tollet2016, Elzant2016}, while for massive galaxies, shallower DM density slopes require more energetic events. 
Supermassive black hole (SMBH) feedback is considered effective at generating outflows and affecting baryonic and DM properties in massive galaxies.
The high energy output of AGN feedback makes DM density profiles less cuspy for massive halos \citep{Peirani2008, Martizzi2013, Peirani2017, Maccio2020, Dekel2021, Jahn2023}.
Furthermore, the energetics of AGN feedback can also make DM halos in massive galaxies more prolate \citep{Nunez2023, Naree2023}.
Indeed, the energy injected by these feedback events (both stellar and AGN) evacuates baryons from the central regions causing rapid fluctuations in the local gravitational potential.
Such variations in the gravitational potential alter the orbits of DM particles leading to changes in the inner DM content in galaxies at all halo mass scales \citep{Blumenthal1986, Peirani2008, Governato2010, Pontzen2012, Peirani2017, Lovell2018}.

This paper reports the impact of SMBH on the central DM distribution in galaxies using the Numerical Investigation of a Hundred Astrophysical Object (NIHAO) project \citep{nihao_main, Blank2019}.
In particular, the amount of DM mass which is displaced due to AGN feedback is quantified, and the spatial scale over which the DM displacement is observed also characterized.
NIHAO simulations have already proven successful at matching various observational aspects of galaxy formation and evolution such as the local galaxy velocity function \citep{Maccio2016}, high-redshift clumps in galaxy discs \citep{Buck2017}, the baryonic Tully-Fisher relation \citep{Dutton2007, Arora2023}, properties of low surface brightness galaxies \citep{dicintio2019}, the presence (or lack thereof) of diversity in galaxy rotation curves \citep{santos2018, Frosst2022}, the star formation main sequence \citep{Blank2021, Blank2022} and various structural scaling relations \citep{Arora2023}.
More details about these simulations are found in \sec{nihao}. 

This paper is organized as follows: 
\sec{nihao} describes the set of NIHAO zoom-in hydrodynamical simulations, with and without AGN sub-grid prescriptions.
The amount of DM suppression as a function of radius and time, as well as the evolution of individual galaxy components (star, gas and DM) with time, are addressed in Sections~\ref{sec:dm_deficit} and \ref{sec:mass_evol}, respectively.
The spatial extent of the DM suppression due to AGN feedback and its evolution with time are discussed in \sec{bh_influence}. 
We conclude in \sec{conclusion} with a summary of the impact of SMBH feedback on the DM halos of intermediate and high-mass galaxies. 

\section{NIHAO galaxy formation simulations}
\label{sec:nihao}

We have used the Numerical Investigation of a Hundred Astrophysical Objects (NIHAO; \citealt{nihao_main, Blank2019}) project to study the impact of AGN feedback on the DM distribution in the inner parts of galaxies.
The NIHAO project incorporates $\sim$130 cosmological zoom-in simulations of galaxies with stellar masses ranging from $10^{6}-10^{12}\,{\rm M_{\odot}}$ at \rz.
These simulations were performed with a flat $\Lambda\rm{CDM}$ cosmology parameter from the \cite{planck14}, using TreeSPH code {\scriptsize GASOLINE2.0} \citep{Wadsley2004, Wadsley2017}.
All simulations have similar DM resolution, containing approximately $10^6$ dark particles with a softening length of $\epsilon_{\rm dark} = 931\,{\rm pc}$ and mass resolution of $1.7\times 10^6\,{\rm M_{\odot}}$.
Furthermore, each simulation consists of approximately $10^6$ gas particles with a softening length of $\epsilon_{\rm gas} = 397\,{\rm pc}$ and a typical particle mass of $3.1\times 10^5\,{\rm M_{\odot}}$.

The NIHAO galaxies were allowed to form stars according to the Kennicutt-Schmidt law \citep{Kennicutt1998,Sun2023} with suitable temperature and density thresholds, $\rm T<15000\,K$ and $\rm n_{th}>10.3\, cm^{-3}$. 
The rate of star formation is characterized as $\dot{M}_{*}=c_*M_{\rm gas}t_{\rm dyn}^{-1}$, where $t_{\rm dyn}=(4\pi G\rho)^{-1/2}$ is the gas particle's dynamical time which has a mass of $M_{\rm gas}$ and a density, $\rho$.
The star formation happen with a star formation efficiency of $c_* =0.1$.
Energy is re-injected into the interstellar medium (ISM) from stars through blast-wave supernova feedback. 
Massive stars also ionize the ISM before their supernova explosion~\citep{Stinson2006, nihao_main}. 
This ``early stellar feedback" (ESF) mode is set to inject 13 per cent of the total stellar flux of $2\times 10^{50}\,{\rm erg\,M_{\odot}^{-1}}$ into the ISM. 
This differs from the original prescription of \cite{Stinson2013} to account for increased mixing and conform to basic abundance matching prescriptions 
\citep{Behroozi2013}. 
For the supernova feedback, massive stars with $\rm 8\,M_{\odot}<M_{star}<40\,M_{\odot}$ inject energy and metals into the ISM. 
This energy is injected into high density gas and radiated away due to efficient numerical cooling \citep{Stinson2006}. 
For gas particles inside the blast radius, cooling is delayed by $30\,{\rm Myr}$ \citep{Stinson2013}.
The stellar feedback does not have any variability with halo mass and/or redshift.

SMBHs (AGN) and their associated feedback were also included in the revised NIHAO simulations \citep{Blank2019}.
All NIHAO galaxies with halo mass $M_{200}>5\times 10^{10}\,\rm {M_{\odot}}$, where $M_{200}$ is the total mass within the radius $R_{\rm 200}$ (where the mass density achieves a value of 200 times the critical density of the Universe) were seeded with a SMBH of mass $M_{\rm BH,i}\sim 10^{5}\,\rm{M_{\odot}}$ (corresponding the mass of a gas particle).
Seeding is performed by converting a gas particle with the lowest 
gravitational potential to a black hole particle with mass $M_{\rm BH,i}$.
In some cases, the mass of the gas particle used for seeding may be less (due to star formation) than the seed mass stated above.
To compensate for stochastic motion of the BH due to accretion or dynamical friction which displaces it from the gravitational potential minimum, the position and velocity of the BH particle is set at each time step to the DM particle with the lowest gravitational potential within 10 softening lengths.
SMBH accretion within the Bondi-Hoyle-Lyttleton parametrization \citep{Bondi1952} employs the mass of the BH ($M_{\rm BH}$) and the density ($\rho$), sound speed ($c_{\rm s}$) and velocity ($v$) of the surrounding gas as
\begin{equation}
    \dot{M}_{\rm BHL} = \frac{4\pi\alpha G^2 M_{\rm BH}^2\rho}{(c_{\rm s}^2+v^2)^{3/2}}.
\end{equation}
In the equation above, $G$ is the gravitational constant and the parameter $\alpha=70$ accounts for the numerical resolution of the simulation. 
The accretion of the SMBH is limited by the Eddington rate ($\dot{M}_{\rm Edd}=M_{\rm BH}/\tau_{\rm s}\epsilon_{\rm r}$) where $\tau_{\rm s}=4.5\times 10^{8}\,{\rm yr}$ is the Salpeter time-scale \citep{Salpeter1964} and $\epsilon_{\rm r}=0.1$ is the radiative efficiency \citep{Shakura1973}.
The accretion of SMBH in NIHAO galaxies is then defined as 
\begin{equation}
    \dot{M}_{\rm BH} = min(\dot{M}_{\rm BHL}, \dot{M}_{\rm Edd}).
\end{equation}
The accretion of the SMBH results in a luminosity which is used as feedback with a feedback efficiency of $\epsilon_{\rm f}=0.05$.
This fraction of the SMBH luminosity is distributed to the 50 nearest gas particles. 
To prevent large sound speeds the specific energy of a single gas particle is limited to $(0.1c)^2$.
Further details on the AGN implementation in NIHAO are presented in \cite{Blank2019} and \cite{Soliman2023}.

NIHAO galaxies simulated according to this scheme have been shown to match various observed galaxy properties and scaling relations (\sec{Intro}), including the black hole mass - stellar mass, $M_{\rm BH}-M_{*}$, relation (see next Section and \Fig{mstar_mbh}).
For this study, we used an extensive version of the sample presented in \cite{Blank2019}; in particular, we have added the AGN version of the low-mass classic NIHAO galaxies \citep[presented in][]{nihao_main} to the simulated sample presented in \cite{Blank2019}.

\subsection{Sample Selection}
\begin{table*}
\begin{tabular}{@{}ccccc@{}}
\toprule
Bin               & Stellar Mass                                      & N   & Mean [$\log_{\rm 10}(M_*/M_{\odot})$] & Standard Deviation \\ 
(1)               & (2)                                               & (3) & (4)                                       & (5)                \\ \midrule
Low-mass          & $9.5\leq\log_{\rm 10}(M_*^0/{\rm M_{\odot}})<10.0$  & 17  & \phantom{0}9.66                                      & 0.20               \\
Intermediate-mass & $10.0\leq\log_{\rm 10}(M_*^0/{\rm M_{\odot}})<10.5$ & 14  & 10.24                                     & 0.13               \\
High-mass         & $10.5\leq\log_{\rm 10}(M_*^0/{\rm M_{\odot}})<11.0$ & 15  & 10.83                                     & 0.16               \\ \bottomrule
\end{tabular}
\caption{Stellar mass bins for NIHAO galaxies used in this study.
Columns (1-2) give the stellar mass bin and their respective stellar mass limits. 
Columns (3-5) report the number of bins, the mean and the standard deviation in stellar mass in each bin respectively. 
The binning is performed using stellar mass at $z=0$ measured as the sum of all stellar particles within a sphere of radius $0.2R_{\rm 200}$.}
\label{tab:mstar_bins}
\end{table*}

NIHAO enables galaxy simulations with a range of feedback mechanisms (e.g., AGN vs. stellar). 
While the classic NIHAO simulations \citep{nihao_main} only included stellar feedback (SF) for all galaxies, the revised version of NIHAO simulations \citep{Blank2019} also treated SMBH feedback.
Given our goal to isolate the coupling of DM and AGN feedback in galaxies, we compared the \noagn and \agn simulations over time.
For this study, we used galaxies with 
$9.5\leq\log(M_*/{\rm M_{\odot}})<11.0$ from the \agn simulations.
The simulated \agn galaxies are binned using $z=0$ stellar mass measurements (within $0.2R_{\rm 200}$) described in \Table{mstar_bins}.
These stellar bins are used for the remainder of this study.
The \noagn counterparts are the same \agn galaxies simulated without AGN feeedback such that the only feedback mechanism comes from the stellar processes.
Dark Matter Only (DMO) NIHAO simulations for the \agn sample were also used for comparison (see \sec{mass_evol}). 
For all galaxies, we have used distribution masses (dark, star, gas, and cold gas) as a function of radius. 
The regions specified by these radii are spherical such that projection effects are nullified.

\begin{figure*}
    \centering
    \includegraphics[width=\linewidth]{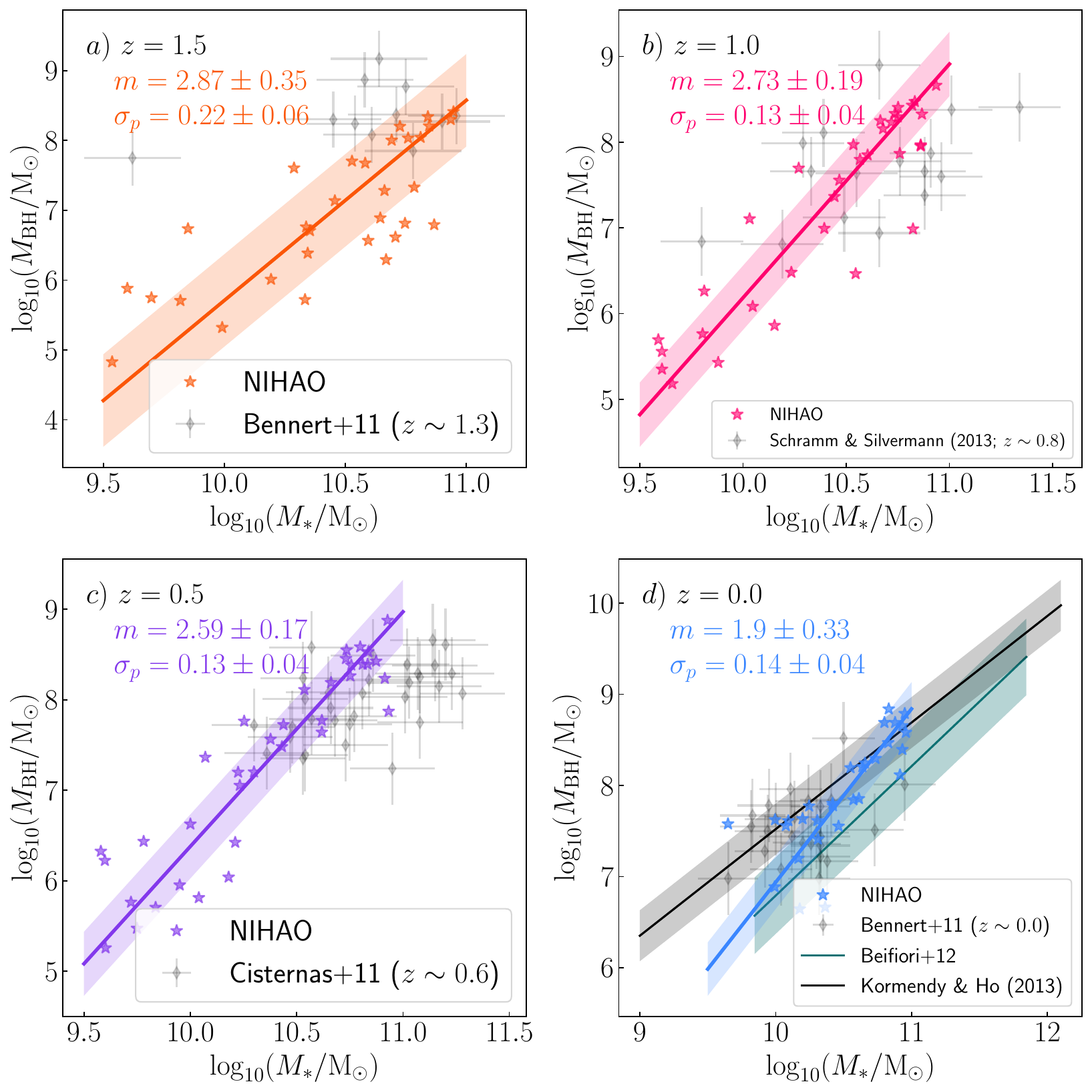}
    \caption{Black hole mass vs. stellar mass relations at four redshifts (see top left corner of each panel) for NIHAO galaxies.
    The solid line and the shaded regions represent the orthogonal best fit and the orthogonal scatter for the relations, respectively. 
    Observational comparisons from various literature sources listed in the legend for the appropriate redshift.}
    \label{fig:mstar_mbh}
\end{figure*}

To validate that NIHAO simulations produce realistic galaxies, we first present central black hole mass -- stellar mass ($M_{\rm BH}-M_{*}$) relations, along with their slope and orthogonal scatter, at four different redshifts in \Fig{mstar_mbh}.
All simulations are seen to follow a linear $M_{\rm BH}-M_{*}$ relation at all redshifts.
With increasing age of the Universe, from past to present, the slope of the $M_{\rm BH}-M_{*}$ relation decreases while the scatter evolution is approximately constant for $z<1$.
The best fit relations shown in \Fig{mstar_mbh} differ from those of \cite{Blank2019} in light of new simulated low mass galaxies with $\log(M_*/{\rm M_{\odot}})<10.0$. 

\Fig{mstar_mbh} also features observational results (grey diamonds with errorbars) at all redshifts by \cite{Bennert2011b}, \cite{Bennert2011}, \cite{Cisternas2011}, and \cite{Schramm2013}, with masses re-calibrated using the prescription of \cite{Ding2020}.
Reassuringly, the observed samples show broad agreement with NIHAO galaxies at all redshifts. 
However, this agreement only holds for high stellar mass galaxies due to selection effects in the observations, where observed low stellar mass galaxies, hosting low-mass SMBH, are limited at high-redshifts.
At $z=0$, the NIHAO galaxies are also compared with the observed relations (solid line and shaded region) of \cite{Beifiori2012} and \cite{Kormendy2013}.
Broadly, the simulated NIHAO galaxies at present day occupy the same parameter space as observed by \cite{Bennert2011}, \cite{Beifiori2012} and \cite{Kormendy2013}.
This data-model comparison reaffirms NIHAO's ability to match the observed Universe.

\section{Dark Matter Supression} \label{sec:dm_deficit}

\begin{figure}
    \includegraphics[width=\columnwidth]{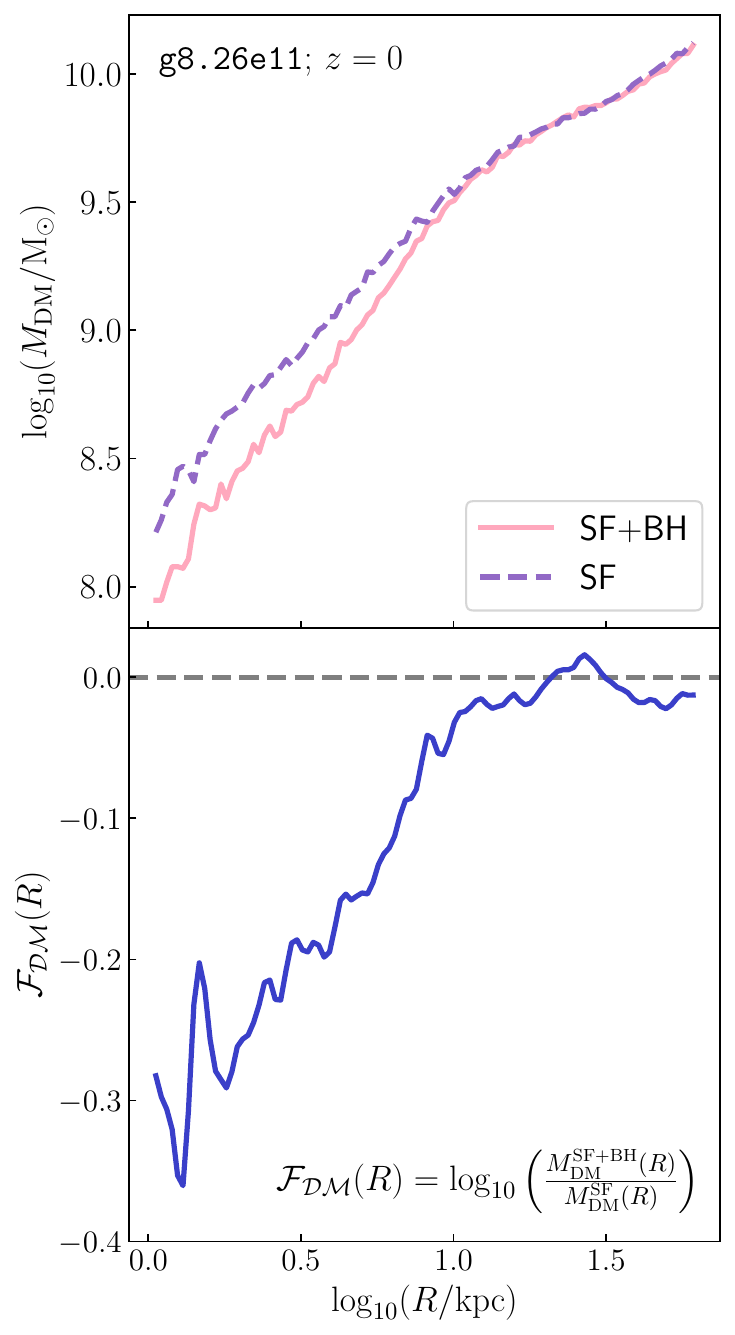}
    \caption{DM properties for NIHAO simulated \texttt{g8.26e11} galaxy. 
    The top panel shows the dark matter mass profiles for \agn (solid pink) and \noagn (dashed purple) simulations at $z=0$. 
    Masses are measured in 100 spherical shells logarithmically spaced between $0.005R_{\rm 200}$ and $0.3R_{\rm 200}$.
    The bottom panel shows the DM suppression (\fdm) as a function of radius for the same galaxy. 
    The inset text gives the equation for \fdm. 
    The grey horizontal line shows \fdm$=0.$ corresponding to no effect of AGN on the DM halo.}
    \label{fig:fdm_calc}
\end{figure}

To study the halo response due to AGN feedback on NIHAO galaxies, we compare the \agn and \noagn simulations. 
The DM suppression due to AGN feedback is calculated as the ratio of DM mass between \agn and \noagn simulations at a given radius. 
Firstly, the halo of interest is rotated and re-positioned in the x-y plane such that the net angular momentum vector is aligned parallel to the z-axis.
We use the \texttt{pynbody.analysis.profile} submodule and measure DM mass within 100 spherical shells logarithmically spaced between $0.005R_{\rm 200}$ and $0.3R_{\rm 200}$.
Then, the DM suppression, \fdm(R), at a radius, $R$, is defined as 
\begin{equation}
    \mathcal{F_{DM}}(R) = \log_{10}\Bigg(\frac{M_{DM}^{SF+BH}(R)}{M_{DM}^{SF}(R)}\Bigg),
    \label{eq:fdm} 
\end{equation}
where a thin shell is centered around around R, $M_{DM}^{SF+BH}(R)$ is the DM mass in the shell for the \agn simulations, and $M_{DM}^{SF}(R)$ is the same quantity in the \noagn simulations.
A graphical representation for the DM and \fdm profiles for a sample NIHAO galaxy (\texttt{g8.26e11}) is presented in \Fig{fdm_calc}.
This procedure results in a DM suppression profile for all galaxies at various redshifts.
In this formalism, $\mathcal{F_{DM}}(R)\sim 0.0$ shows (particularly for small radii) negligible contribution of black hole feedback in removing DM.

\begin{figure*}
    \includegraphics[width=\linewidth]{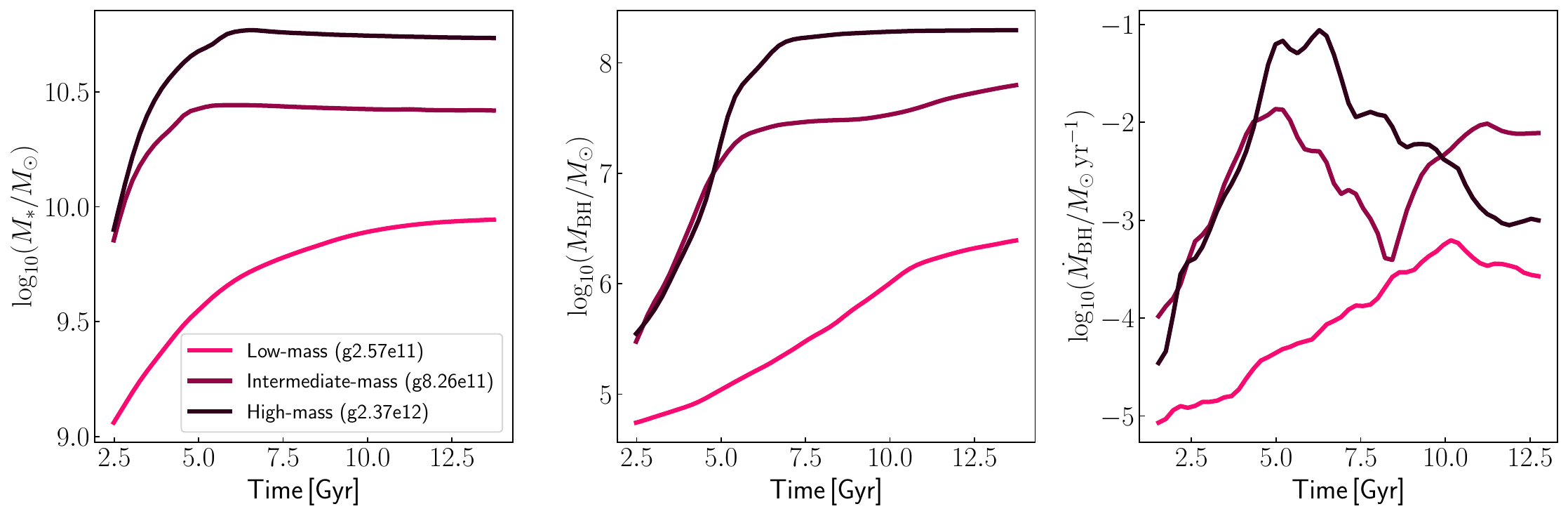}\\
    \includegraphics[width=\linewidth]{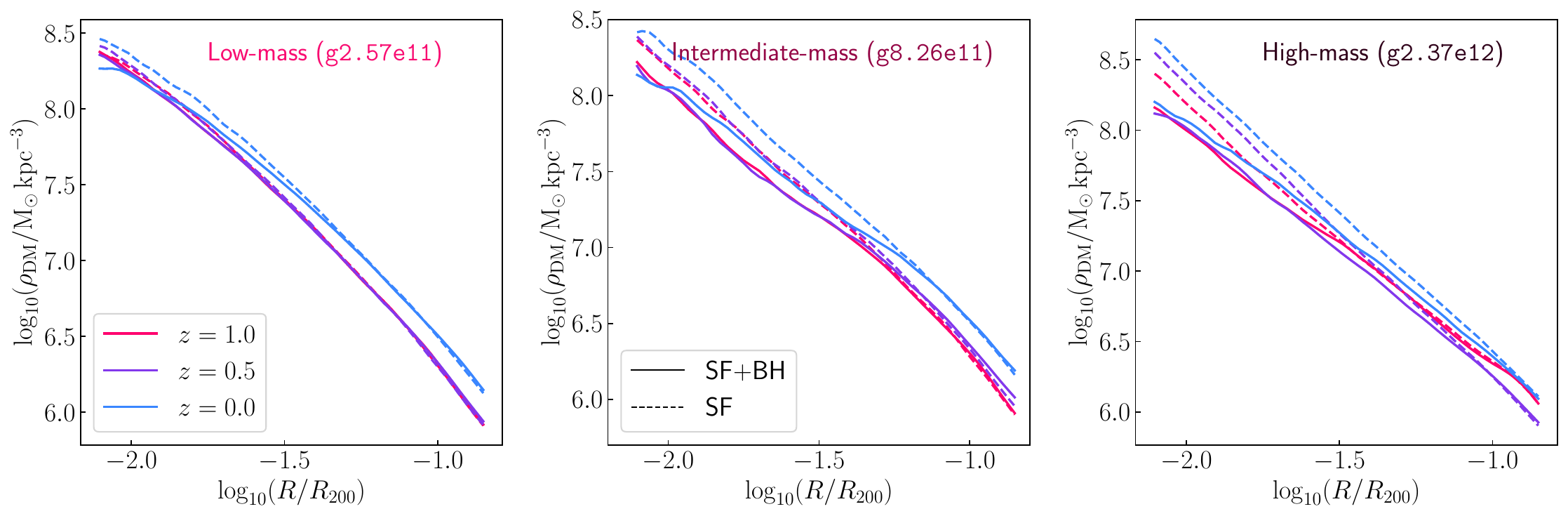}
    \caption{Top row: Evolution of stellar mass ($\log_{\rm 10}(M_*/M_{\rm\odot})$; left-hand panel), central black hole mass ($\log_{\rm 10}(M_{\rm BH}/M_{\rm\odot})$; middle panel), and black hole accretion rate ($\log_{\rm 10}(\dot{M}_{\rm BH}/M_{\rm\odot}{\rm yr^{-1}})$; right-hand panel) as a function of time.
    The SMBH accretion rates presented here are averaged over a time period of 217 Myr.
    Each colour corresponds to a different NIHAO simulation galaxy which belongs to the different stellar mass bin; \texttt{g2.57e11} -- low-mass, \texttt{g8.26e11} -- intermediate-mass, and \texttt{g2.37e12} -- high-mass bins (see \Table{mstar_bins}).
    Bottom row: Dark matter density profiles at $z=1.0, 0.5, 0.0$ (shown as different colors) for the \agn and \noagn simulations presented as solid and dashed lines respectively. 
    Each panel corresponds to a different NIHAO simulation galaxy belonging to a different stellar mass bin; \texttt{g2.57e11} -- low-mass (left-hand panel), \texttt{g8.26e11} -- intermediate-mass (middle pane), and \texttt{g2.37e12} -- high-mass (right-hand panel) bins.}
    \label{fig:single_gal}
\end{figure*}

The top row of \Fig{single_gal} shows the evolution of stellar and SMBH masses, and the SMBH accretion rate with time for typical galaxies within each stellar mass bin described in \Table{mstar_bins}.
As expected for all three galaxies, $M_{*}$ (left-hand panel) and $M_{\rm BH}$ (middle panel) almost monotonically increase to the present day. 
The most massive central black hole is found in the most massive (\texttt{g2.37e12}) NIHAO system.
The SMBH accretion rates presented in the right-hand panel of \Fig{single_gal} are time-averaged over 270\,Myr.
Both intermediate-mass and high mass NIHAO galaxies have peak $\dot{M}_{\rm BH}$ at $t\sim 5\,{\rm Gyr}$.
Conversely, for the low-mass galaxy, the peak accretion rate is found at $t\sim 10\,{\rm Gyr}$.

The impact of AGN accretion and feedback is also imprinted on the central dark matter content. 
The bottom row of \Fig{single_gal} compares the central dark matter density profiles for the \noagn and \agn simulations at redshifts of 1.0, 0.5 and 0.
For $z>0.5$, the DM profiles show no difference between the \agn and \noagn for low-mass NIHAO galaxy.
While the presence of SMBH does not affect the DM distribution at high redshift for the low mass simulation, the rising SMBH mass and accretion rate creates an imprint on the DM profile at $z=0$.
Both intermediate- and high-mass galaxies present different central DM profiles as a result of the AGN accretion and feedback. 
While the central DM densities in the \noagn simulations increase with time, feedback from the SMBH can regulate the DM densities over the evolutionary history; presenting a balance between the halo contraction (due to baryon accretion) and expansion (due to feedback). 
Furthermore, the spatial imprint of DM suppression between the \agn and \noagn simulations increases with  stellar mass bin; the high-mass bin galaxies have the largest radius where suppression differences are found.
Indeed, the larger SMBH and accretion rates can remove more dark matter from the central parts of the intermediate- and high-mass NIHAO galaxies.

\begin{figure}
    \includegraphics[width=\columnwidth]{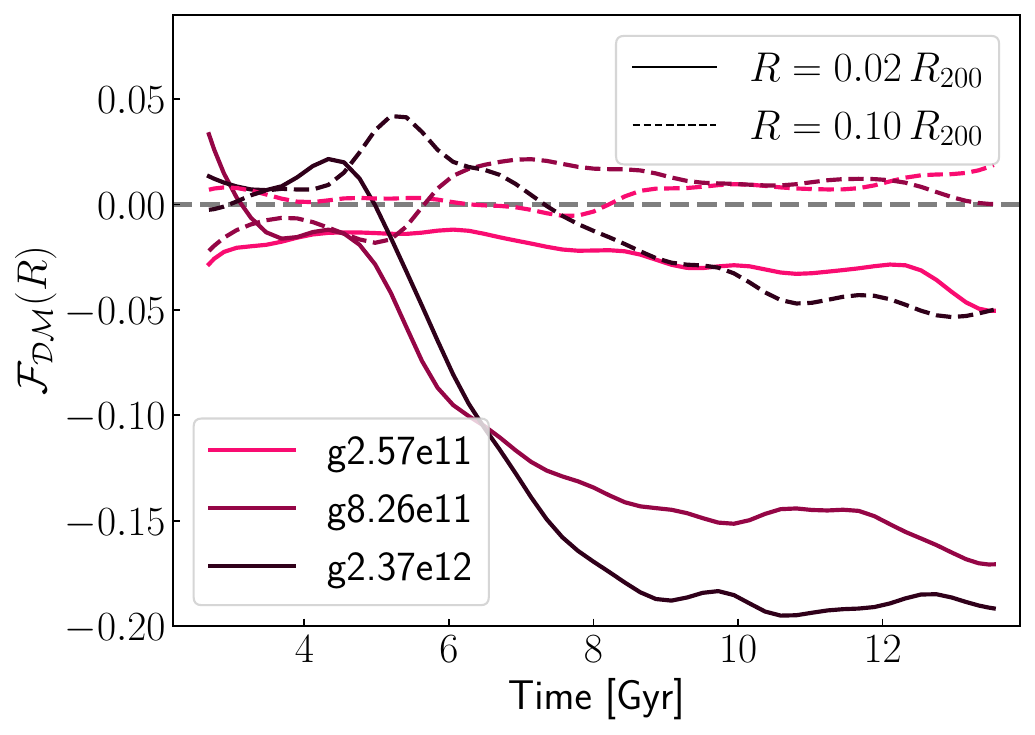}
    \caption{\fdm as function of time (in Gyr) for the three galaxies presented in \Fig{single_gal}. Each colour corresponds to a different NIHAO simulation galaxy which belongs to the different stellar mass bin; g2.57e11 – low-mass, g8.26e11 – intermediate-mass, and g2.37e12 – high-mass bins (see \Table{mstar_bins}). The DM suppression is presented for two radii $R=0.02\,R_{\rm 200}$ (solid line) and $R=0.10\,R_{\rm 200}$ (dashed line). The grey dashed horizontal line shows \fdm$=0$ epresenting no DM deficit due to AGN feedback.}
    \label{fig:single_gal_fdm}
\end{figure}

Figure \ref{fig:single_gal_fdm} illustrates the variations of \fdm at $R=0.02R_{\rm 200}$ and $R=0.10R_{\rm 200}$ with time. 
In cases where a small central SMBH exists, such as \texttt{g2.57e11}, the system does not experience a substantial DM suppression. 
Conversely, within the high-mass bin, \texttt{g2.37e12} exhibits a growing DM suppression, reaching a value of \fdm$\sim -0.2$ at 2 per\,cent of R$_{\rm 200}$ in the present day.
The trend of higher \fdm values at smaller radii can be attributed to the influence of the SMBH on the central gravitational potential. 
Additionally, the \fdm as a function of redshift exhibits an oscillatory behavior, which is closely tied to the active and quiescent periods of AGN feedback (see right-hand top \Fig{single_gal}). 
The outflows generated by AGN feedback lead to alterations in the local gravitational potential, resulting in an increased dark matter deficit. 
For $z\lesssim 1.5$, both galaxies in intermediate- and high-mass bins rapidly decrease in \fdm which coincides with the sharp increase in central SMBH mass.
At some redshifts, the values of \fdm measured at $R=0.10R_{\rm 200}$ show a DM excess relative to \noagn simulations due to merger activity.

\begin{figure*}
    \includegraphics[width=\linewidth]{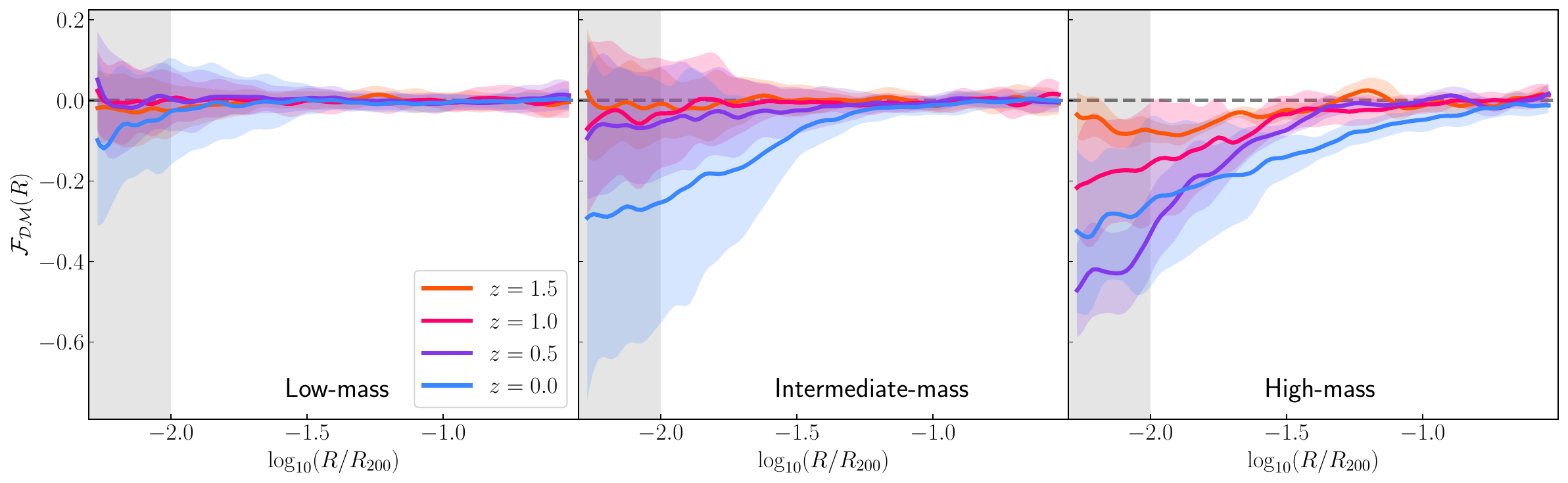}
    \caption{Radial profiles of DM suppression as defined in \Eq{fdm}, at four different redshifts (see text inset) in three stellar mass bins (\Table{mstar_bins}) for NIHAO galaxies; left-hand panel -- low-mass, middle panel -- intermediate-mass, and right-hand panel -- high-mass bin respectively. 
    The solid thick line and coloured shaded region show the median profiles and $1\sigma$ scatter of the distributions at different redshifts, respectively. 
    In all panels, the grey shaded region represents the DM resolution limit ($<0.01\,R_{200}$).}
    \label{fig:dm_ratio}
\end{figure*}

\Fig{dm_ratio} shows $\mathcal{F_{DM}}$ as a function of $R/R_{\rm 200}$ for NIHAO galaxies at four different redshifts (from $z=1.5$ to present day).
The DM suppression profiles are binned using the present day stellar mass (see \Table{mstar_bins}) and DM suppression profile medians are calculated as a function of radius (solid lines in \Fig{dm_ratio}).
The spread of the distribution, 1$\sigma$ scatter range, is highlighted by the shaded region in \Fig{dm_ratio}.

Throughout their evolutionary history, low-mass NIHAO systems have a median $\mathcal{F_{DM}}\sim 0$ indicating negligible impact of black hole feedback on the DM halo.
This is indeed expected as low mass galaxies population host mostly low-mass central black holes $\log(M_{BH}/M_{\odot})\lesssim 5.5)$ with minimal AGN energetics.
Indeed, low-mass NIHAO galaxies show little SMBH mass evolution beyond the black hole seed mass.  
At $z=0.5$ and $z=0.0$, $\mathcal{F_{DM}}\sim -0.1\,{\rm dex}$, low-mass galaxies show a small DM suppression albeit within the resolution limits of the simulations (${\sim}0.01\,R_{\rm 200}$ presented as the shaded grey regions in all panels of \Fig{dm_ratio}).
A number of low mass galaxies have either a DM excess or a deficit at larger radii. 
Closer examination of these galaxies shows evidence of either major or minor companion subhalos within $R\sim 0.3\,R_{\rm 200}$ leading eventually to a merger.
These small deficits are largely indicative of interactions and cosmic stochasticity caused by running the same simulation multiple times (however with different hydrodynamic prescriptions). 
This is demonstrated by the scatter (shaded region in the left-hand panel of the \Fig{dm_ratio}) of low mass NIHAO galaxies at $z=0.0$ distributed symmetrically around $\mathcal{F_{DM}}=0$.

The middle panel of \Fig{dm_ratio} shows Milky Way mass spirals from the NIHAO simulations at four redshifts.
For $z<0.5$, all intermediate mass galaxies have systematically \fdm$<0.0$ well outside the resolution limit of the simulations; with a significant \fdm recorded at present day.
Furthermore in the inner parts, in comparison to the low-mass bin, the amplitude of \fdm increases for the intermediate mass galaxies.
With the increasing stellar (halo) mass, the central black hole masses are expected to rise which amplifies the AGN energetics and thus the DM deficits.
While the latter are largely absent at $z=1.5$ ($t_{\rm Universe}\sim 4\,{\rm Gyr}$), they are present at $z=0.5$ and keep on increasing with time.
By present day ($z=0.0$), \fdm$\sim -0.2$ as a result of AGN feedback and the deficit persists out to $\sim$5 per\,cent of R$_{\rm 200}$. 
Intermediate mass NIHAO galaxies host central black holes with 
$\log (M_{\rm BH}/M_{\odot})\sim 7.5$ with energetic AGN feedback capable of altering the inner DM profile \citep{Blank2022}, by removing $\sim 40$ per cent of the DM mass relative to \noagn simulations.

With more massive central black holes, the high-mass bin of NIHAO galaxies (right-hand panel of \Fig{dm_ratio}) shows further DM mass deficits relative to the \noagn simulations.
Significant DM deficits for the high-mass bins are conspicuous at higher redshifts (compared to lower mass cases) given their more massive central black holes.
While \fdm$\sim0.0$ for the intermediate mass systems at $z=1.5$; the high-mass galaxies have \fdm$<0.0$. 
Furthermore, the DM deficits grow in amplitude and expand radially as redshifts approach $z=0$.
At $z=0.0$, the DM deficits in the \agn simulations grow to \fdm$\sim -0.2$ in the inner parts, similar to the ones for intermediate mass galaxies at present day.
Interestingly, the maximum \fdm for high mass galaxies is found at $z=0.5$ with high mass halos showing signatures of contraction and/or dynamical relaxation. 
The increase in DM content from $z=0.5$ to $z=0.0$ seems to be linked to the evolution of the SMBH mass in the high-mass bin.

\begin{figure*}
    \includegraphics[width=\linewidth]{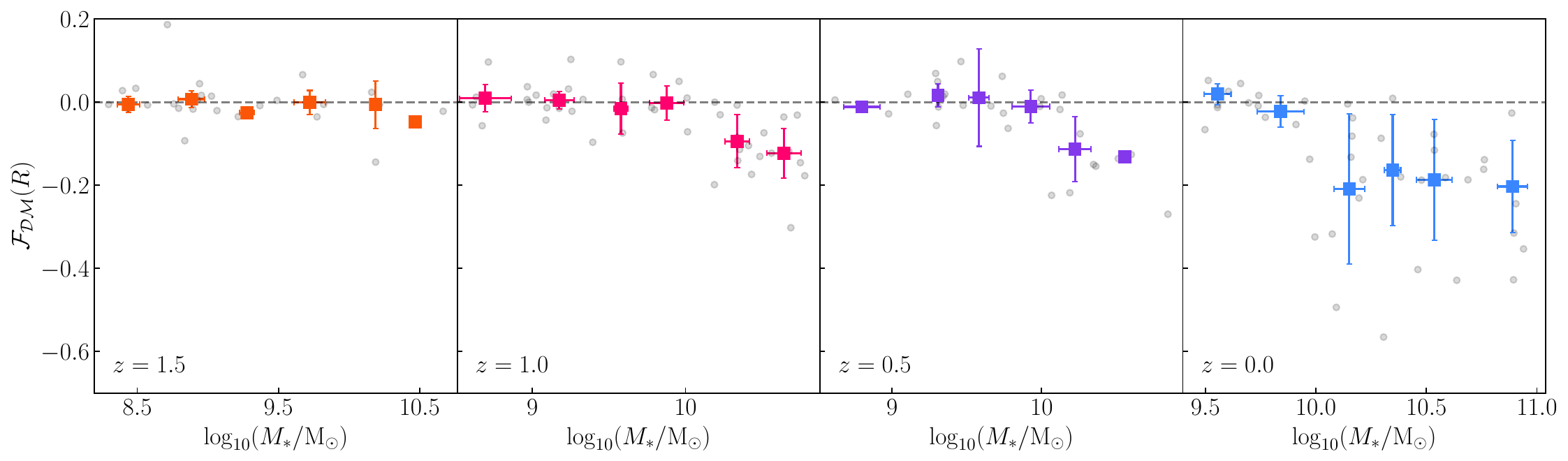}
    \caption{\fdm$(R)$ versus stellar mass for all NIHAO galaxies at different redshifts, evaluated at $0.02\,R_{\rm 200}$. 
    The grey points show individual NIHAO systems while the coloured squares and error bars represent the median \fdm and its scatter within a stellar mass bin. 
    At each redshift, all galaxies are binned into six stellar mass bins.}
    \label{fig:dm_ratio_mstar}
\end{figure*}

\Fig{dm_ratio_mstar} shows variations of DM suppression versus stellar mass at $R=0.02R_{\rm 200}$, with data for the individual NIHAO galaxies and median trends at various redshifts.
At each redshift, the DM deficits increase by $\sim -0.2$ dex with increasing stellar mass (from $\log_{10}(M_*/M_{\rm\odot})=9.5-11.0$).
For high-mass systems, the DM suppression shows a pronounced increase as well, albeit with larger scatter.
While DM deficits are negligible at $z=1.5$, the average \fdm grows to $\sim$-0.2 by present day.
Systems at $z<1$ and $\log_{10}(M_*/{\rm M_{\odot}})\gtrsim 10$ have approximately constant \fdm, albeit based on low-number statistics. 
However, the constant trend of DM suppression with stellar mass is magnified at $z=0$ where DM deficits for NIHAO systems with $\log_{\rm 10}(M_*/M_{\odot})>10$
averages at \fdm$\sim -0.2$.
At all redshifts, galaxies were also found with a DM excess in \agn simulations relative to \noagn simulations due to recent merger activity.

The amount of DM deficits is expected to correlate to the mass of the central black hole (or integrated black hole luminosity) and the gas availability for accretion/ejection onto the SMBH that alters the local gravitational potential. 
Therefore, the constant \fdm for large stellar masses informs us about the AGN feedback coupling with DM content in the central parts.
The next section addresses the mass evolution of different galaxy components to explain the constant trend of the DM deficits. 

\section{Mass Evolution}\label{sec:mass_evol}

We now study the mass evolution of various galactic components in an attempt to better understand their impact on the DM suppression in the central regions.
We first examine the evolution of the central black hole followed by the baryonic contents and finally the DM mass and fraction. 
Unless otherwise stated, the three stellar mass bins presented in \Fig{dm_ratio} are preserved for the remaining analysis.

\subsection{Black Hole}

\begin{figure*}
    \centering
    \includegraphics[width=\linewidth]{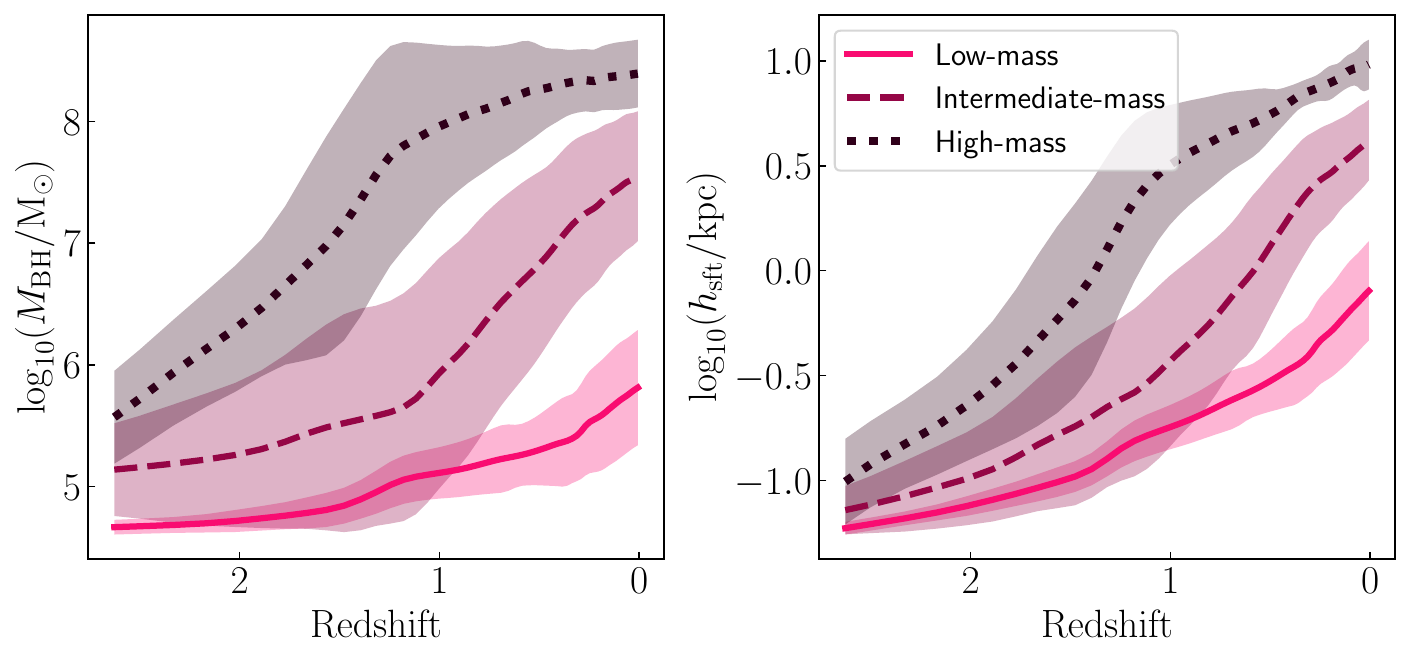}
    \caption{Evolution of the central black hole mass (left-hand panel) and softening length (right-hand panel) for the \agn simulations. 
    The solid, dashed and dotted lines line represent respectively 
    the low-mass, intermediate-mass and high-mass systems (\Table{mstar_bins}). 
    The lines represent the median of each stellar mass bin while the shaded region corresponds to the $1\sigma$ region.}
    \label{fig:smbh_evol}
\end{figure*}

\Fig{smbh_evol} shows the evolution of the central black hole for the NIHAO galaxies, in three stellar mass bins. 
For the NIHAO simulations, the central black hole is defined as the most massive black hole particle in the central halo.
The central black holes in three stellar mass bins display rather different behaviours and should therefore affect the DM suppression differently.
For low-mass NIHAO galaxies, the central black holes show negligible growth; only $\sim 1\,$dex from $z=2.5$ to $z=0$.
The low-mass central black holes, and their negligible evolution, would result in minimal feedback leading to the absence of DM deficits in dwarfs galaxies (left-hand panel of \Fig{dm_ratio}), regardless of redshift.

The intermediate-mass galaxies (dashed line and shaded region in \Fig{smbh_evol}) have central black holes which grow monotonically by $\sim$2.5\,dex from $z=2.5$ to $z=0$; however with large scatters.
On average, the mass of the central BH growth presents sharper slopes at $z<1$ for intermediate-mass systems.
Their inner DM deficits also grow between $z=1.5-1$ (\Fig{dm_ratio}).
For $z<1.5$, the inner \fdm grows in lockstep with the central black holes. 
For the present day, median $\log(M_{\rm BH}/M_{\odot})\sim 7.5$ can create an inner DM suppression of ${\sim}-0.2$\,dex.

Finally, the dotted line in \Fig{smbh_evol} represents the SMBH evolution for galaxies in the highest stellar mass bin.
Their SMBHs increase in mass monotonically from $z=2.5 - 1.0$ by a factor $\sim$2 dex.
Interestingly for $z<1.0$, a saturation in the SMBH mass is found, 
with $\log(M_{\rm BH}/M_{\odot})\sim 8.2$ for the remaining time of their evolution. 
The lack of SMBH growth is likely related to the limited gas reservoir in the central parts of such simulated high mass systems. 
The initial SMBH growth results in strong feedback leading to evacuation of the gas from the central regions within those high-mass galaxies. 
The saturation of SMBH mass at late times allows for the DM halos in the \agn simulations to contract yielding a reduction of the DM suppression from \fdm$\sim0.3$ at $z=0.5$ to \fdm$\sim0.25$ at present day.

The right-hand panel in \Fig{smbh_evol} presents the evolution of the numerical softening length for the central SMBH in all three bins. 
Similar to SMBH mass, the softening lengths for the NIHAO galaxies monotonically grow. 
Within NIHAO, the softening lengths of SMBH particles increases by a factor of $(1+\Delta m/m_{\rm BH})^{1/2}$ where $\Delta m$ is the change in SMBH mass.
The measurements of softening length of central SMBH are important quantity as we expect to measure the ``sphere of influence" (see \sec{bh_influence}) for the SMBH on the central DM content.

Next, we address the evolution of baryons (specifically gas) within the inner parts of galaxies and the role of SMBHs on such evolution. 

\subsection{Baryons}

\begin{figure*}
    \centering
    \includegraphics[width=\linewidth]{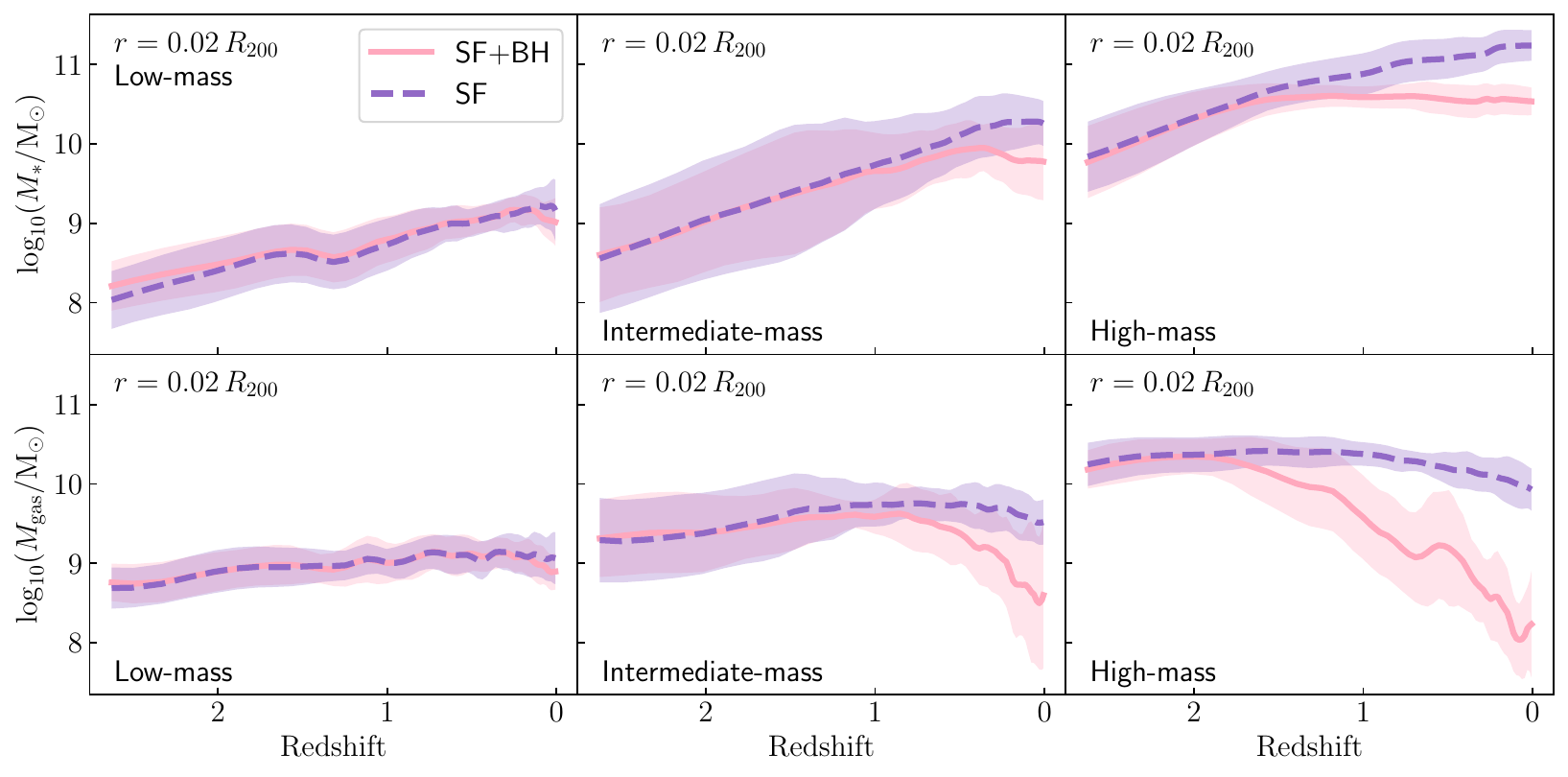}
    \caption{Evolution of the baryonic mass (specifically stellar and gas) for \agn and \noagn simulations within $0.02\,R_{\rm 200}$. 
    The purple-dashed and pink-solid lines correspond to the \noagn and \agn simulations, respectively. 
    Each panel shows a different stellar mass bin (see \Table{mstar_bins}). 
    The lines are the median DM masses in each stellar mass bin while the shaded areas represent the $1\sigma$ region.}
    \label{fig:bar_evol}
\end{figure*}

The presence or absence of AGN feedback certainly affects the baryonic properties of galaxies \citep{Wilman2010, Beifiori2012, Silk2012, Blank2022, Frosst2022, Waterval2022}.
\Fig{bar_evol} shows the stellar (top row) and gas (bottom row) mass as a function of redshift within 2 per\,cent of the virial radius for both \agn and \noagn simulations. 
For low-mass bins (top left-hand panel), the presence of a central SMBH has little impact on the stellar content.
However, both \agn and \noagn simulations in the low-mass bins show a $\sim$1 dex growth in stellar mass.

For all stellar mass bins, the stellar content in the inner regions increases as a function of redshift, with the most growth observed for intermediate-mass systems due to sustained star formation. 
The difference between \agn and \noagn simulations, absent for low-mass systems, is evident for the intermediate- and high-mass bins.
The differences in stellar mass start at $z\sim 1.5$ and increase steadily to the present day.
While \noagn simulations show signatures of overcooling to keep acquiring stellar content in the central parts, the latter is regulated by black hole feedback in the \agn simulations.
The \noagn NIHAO simulations show signatures of overcooling in the high-mass bins yielding higher stellar concentrations than the \agn case at $z=0$ \citep[see][]{Arora2022, Frosst2022, Arora2023}.
The difference in stellar mass is clearest for high-mass bins, where \agn galaxies show no growth in central stellar content since $z\sim 1.5$. 
This leads to a difference of 0.5 dex in the stellar content between the \agn and \noagn high-mass simulated galaxies.

The bottom row of \Fig{bar_evol} shows the evolution of the gas mass for the \agn and \noagn simulations within $0.02\,R_{\rm 200}$.
In the \noagn simulation, for all stellar mass bins, the amount of gas in the central parts remains approximately constant over time. 
Indeed, this gas in \noagn galaxies acts as the fuel for stellar mass build up for all stellar mass bins while accretion from the hot halo keeps the gas mass constant.
On the other hand, the central gas content in the \agn simulations varies with stellar mass. 
While the \agn low-mass bins show identical gas content to the \noagn, the intermediate- and high-mass bins have a smaller amount of gas in the central parts relative to \noagn simulations due to AGN feedback. 
Within the intermediate- and high-mass bins the SMBH can create outflows which empty out the gas content from the central regions.

The gas content between the \agn and \noagn simulations for the intermediate-mass and high-mass bins starts to deviate around $z\sim 1.5-2$.
For the intermediate-mass galaxies, the cold gas content is less in the \agn systems by approximately 1~dex; whereas the difference increases to 2~dex for the high-mass bins.
With more massive black holes (therefore stronger AGN feedback), the central cold gas deficit, relative to \noagn systems, is also two orders of magnitude larger, which explain the lack of stellar build-up in the central parts.
Furthermore, the gas content between the two simulations (\agn and \noagn) deviates from one another at earlier times for the high-mass systems.
The two processes coupled together explain the constant DM suppression for NIHAO galaxies with $\log(M_*/M_{\odot})\gtrsim 10$ (see \Fig{dm_ratio_mstar}).

\subsection{Dark Matter}
\label{sec:dm}

\begin{figure*}
     \centering
    \includegraphics[width=\linewidth]{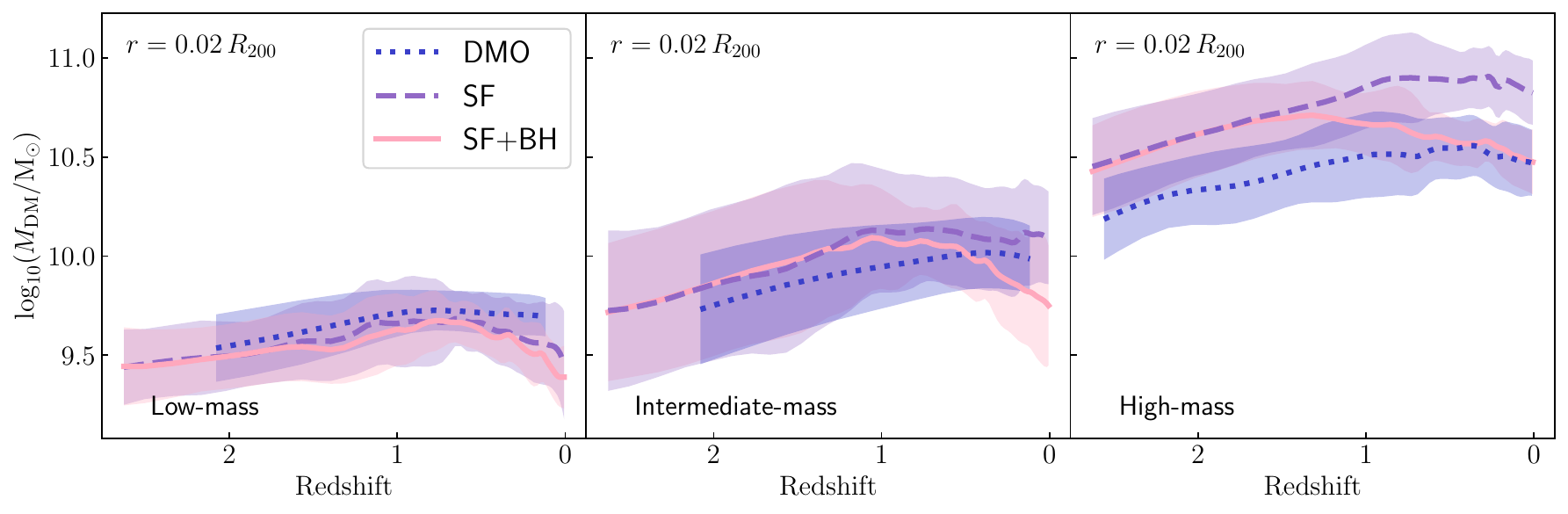}
    \caption{Evolution of the DM mass for the three simulations within $0.02\,R_{\rm 200}$. 
    The blue-dotted, purple-dashed and pink-solid lines correspond to the NIHAO-\textit{DMO}, \noagn and \agn simulations respectively. 
    Each panel represents a different stellar mass bin (see \Table{mstar_bins}). 
    The lines show the median DM mass in a stellar mass bin whereas the shaded areas correspond to the $1\sigma$ region. }
    \label{fig:dm_evol}
\end{figure*}

With the DM suppression quantified, we can calculate the absolute amount of DM in the central parts of both \agn and \noagn simulations. 
To get a better appreciation of the baryonic hydrodynamical processes, we also included DM mass evolution from the DM only (DMO) simulations.
It should be noted that some of the DM suppression presented for high-mass galaxy may be partly associated to non-negligible resolution effects associated to the BH particles.

\Fig{dm_evol} shows the DM mass evolution for the three simulation sets within $0.02\,R_{\rm 200}$.
For low-mass systems in all three simulations, the amount of DM within the central parts increases slightly until $z\sim 1.0$.
For $z<1.0$, the central DM mass in the \agn and \noagn simulations decreases as a result of strong stellar feedback in low-mass galaxies.
Indeed, low mass galaxies within the NIHAO simulations exhibit halo expansions which stems from the competition between gas accretion and outflows from the feedback \citep{Dutton2016}.
However, all three simulations (NIHAO-DMO, \noagn, and \agn) show negligible differences in the amount of DM within 0.02\,R$_{\rm 200}$ as a function of time.

The middle panel of \Fig{dm_evol} shows the central DM mass evolution for intermediate-mass NIHAO galaxies. 
As in low-mass galaxies, the central DM mass in intermediate-mass systems increases until $z\sim 1.0$, decreasing thereafter to the present day.
Indeed, this drop in local DM mass is due to feedback, stellar for \noagn simulations and AGN and stellar feedback for \agn galaxies.
The impact of the AGN feedback reduces the amount DM in the central parts in \agn galaxies, the difference being $\sim 0.4$ dex at present day.
Relative to NIHAO-DMO simulations, intermediate-mass bin galaxies have significantly expanded halos due to strong feedback for most of their history. 
For the \noagn simulations, the deeper gravitational potential and continued star formation continues to contract the dark matter halos leading to similar DM masses in the NIHAO-DMO versions at $z=0$.
However, the presence of AGN and stellar feedback allows for considerable DM halo expansion starting at $z>1.0$.

Similar overall trends in the central DM mass are found for high-mass galaxies. 
DM mass growth is sustained for both \agn and \noagn galaxies until $z>1.0$.
As a result of the AGN feedback, the difference in DM content starts growing at $z>1.5$ rising to a difference of $\sim 0.5$ dex at present day.
As in intermediate-mass galaxies, for most of their evolutionary history both \agn and \noagn galaxies live in more contracted halos relative \citep[see also][]{Martizzi2013} to the NIHAO-DMO simulations.
The presence of AGN feedback expands the \agn halos leading to similar DM mass in the central parts to the NIHAO-DMO galaxies.
Interestingly, any departures in DM content between \agn and \noagn galaxies are delayed relative to the differences in the gas content (see lower right-hand panel of \Fig{bar_evol}).
For the high-mass systems, while the differences in gas content become first apparent at $z\sim 2.0$, the \agn and \noagn DM content is only significant at $z\sim 1.5$.
Indeed, the change of gravitational potential and variations in DM particle orbits occurs with a time delay as a result of the gas ejection due to AGN feedback, in agreement with various n-body simulations \citep{Martizzi2013, Ogiya2022} and cosmological hydro-dynamical simulations \citep{Peirani2017, Espinosa2022}.

\subsection{Dark Matter Fraction}
\label{sec:dmfraction}

\begin{figure*}
    \centering
    \includegraphics[width=\linewidth]{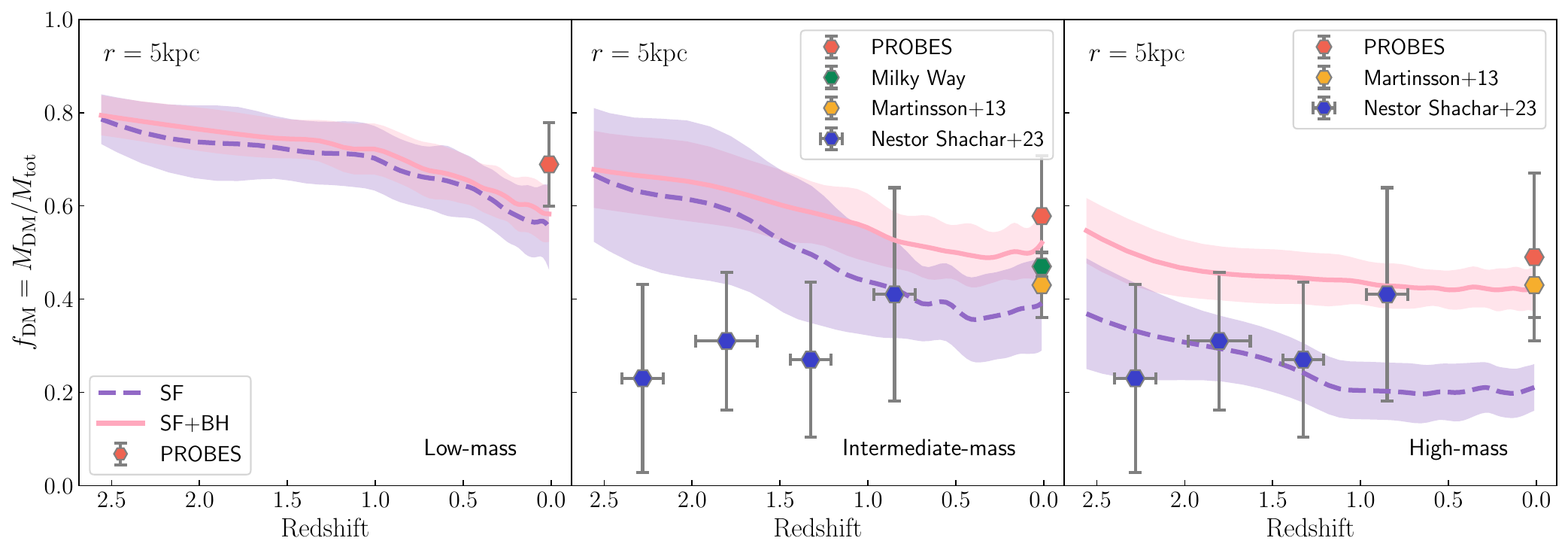}
    \caption{Evolution of the DM fraction for \agn and \noagn simulations. 
    All measurements were made at a physical radius of 5 kpc. 
    The remaining details for this figure are identical to \Fig{bar_evol}.
    Comparisons with $f_{\rm DM}$ inferred from observations are also presented, with measurements from the PROBES catalogue \citep[red hexagon;][]{Stone2022}, various Milky Way studies \citep[green hexagon;][]{Bovy2013, Kafle2014, Piffl2014, Huang2016}, the DISKMASS survey \citep[gold hexagon;][]{Martinsson2013} and the RC100 survey \citep[blue hexagons;][]{Shachar2023}.}
    \label{fig:fdm_evol}
\end{figure*}

In \sec{dm}, we showed that in both intermediate- and high-mass bins, the presence of AGN feedback results in a 0.5 dex DM mass deficit in the central parts (see \Fig{dm_evol}).
We now study the DM suppression signatures on observable quantities such as the relative DM mass fraction, $f_{\rm DM}$, within a radius, R:
\begin{equation}
    f_{\rm DM}(\leq R) = \frac{M_{\rm DM}(\leq R)}{M_{\rm tot}(\leq R)} = \frac{M_{\rm DM}(\leq R)}{M_{\rm DM}(\leq R)+M_{*}(\leq R)+M_{\rm gas}(\leq R)},
\end{equation}
where $M_{\rm DM}$, $M_{*}$, and $M_{\rm gas}$ are the DM, stellar and gas masses within radius, R, respectively.
To enable robust comparisons with observations, we use the radial metric of 5 kpc (proper units) instead of the previously adopted value of 0.02\,R$_{\rm 200}$.

\Fig{fdm_evol} shows the DM fraction within a physical radius of 5\,kpc for \agn and \noagn simulations. 
Regardless of the stellar mass bins, $f_{\rm DM}$ reduces over time for both \noagn and \agn simulations due to an increase in baryonic content and feedback changing dark matter partcles orbits.
In the intermediate- and high-mass regimes, for both simulations, $f_{\rm DM}$ first drops and then levels off beyond $z\sim 1.0$.
For the low-mass systems and both simulations, the $f_{\rm DM}$ evolution is nearly identical, independent of the feedback scenario.
This situation changes for the intermediate- and high-mass bins which have a systematically larger and distinct $f_{\rm DM}$ in the central parts. 
With the addition of a new feedback channel (AGN feedback), the removal of baryons (through outflows and regulating star formation) is more efficient in the \agn simulations than DM removal via gravitational potential fluctuations.
The latter is more prevalent in the high-mass systems, where the \agn simulation yields $\sim$30 per cent larger DM fractions than for \noagn systems.

We have compared the intermediate-mass and high-mass systems with observed estimated DM fractions at different redshifts from \cite{Bovy2013}, \cite{Kafle2014}, \cite{Piffl2014}, \cite{Huang2016}, \cite{Martinsson2013}, and \cite{Shachar2023}.
The high redshift $f_{\rm DM}$ measurements of the RC100 project \citep{Shachar2023} show broad agreement with the \noagn simulations (for $z=2.5-1.0$) in the high-mass regime.
The match between observations and simulations is within the observed errors for the \agn galaxies.
Disagreements also exist between high redshift observations from \cite{Shachar2023} and $f_{\rm DM}$ measurements in the intermediate-mass bin for the \agn simulations.
This is largely associated with the fitting and observation techniques for the RC100 observed data.
In contrast, \cite{Sharma2023} found an approximately constant $f_{\rm DM}\sim 0.6$ over time using 263 galaxies from $z=2.5$ to the present (not shown in \Fig{fdm_evol}), in broad agreement with the intermediate- and high-mass regime of the \agn simulation at higher redshifts.
At $z=0$, the reported average measurements for $f_{\rm DM}$ come from heterogeneous sources (see figure caption) which also show good agreement with NIHAO systems. 
Overall, the \agn simulated galaxies agree well with observations in the intermediate-mass and high-mass regimes. 
It should also be noted that the large errors (0.2\,dex) in $f_{\rm DM}$ estimates largely stem from (uncertain) stellar mass transformations \citep{Conroy2013,Courteau2014}.

Our analysis of the mass evolution of NIHAO simulations has confirmed that the inclusion of SMBH and AGN feedback can affect the DM distribution through through variations of the gravitational potential \citep{Pontzen2012}.
Through star formation and BH accretion, the amount of gas available for SMBH evolution declines rapidly over time, especially for the high-mass NIHAO simulations.
For high-mass systems, this leads to a saturation in the SMBH mass evolution and DM fraction in the central parts for NIHAO galaxies.

Next, we wish to identify the spatial region inside which AGN feedback is most effective at creating a DM suppression.

\section{Sphere of Influence} 
\label{sec:bh_influence}

\begin{figure*}
    \centering
    \includegraphics[width=\linewidth]{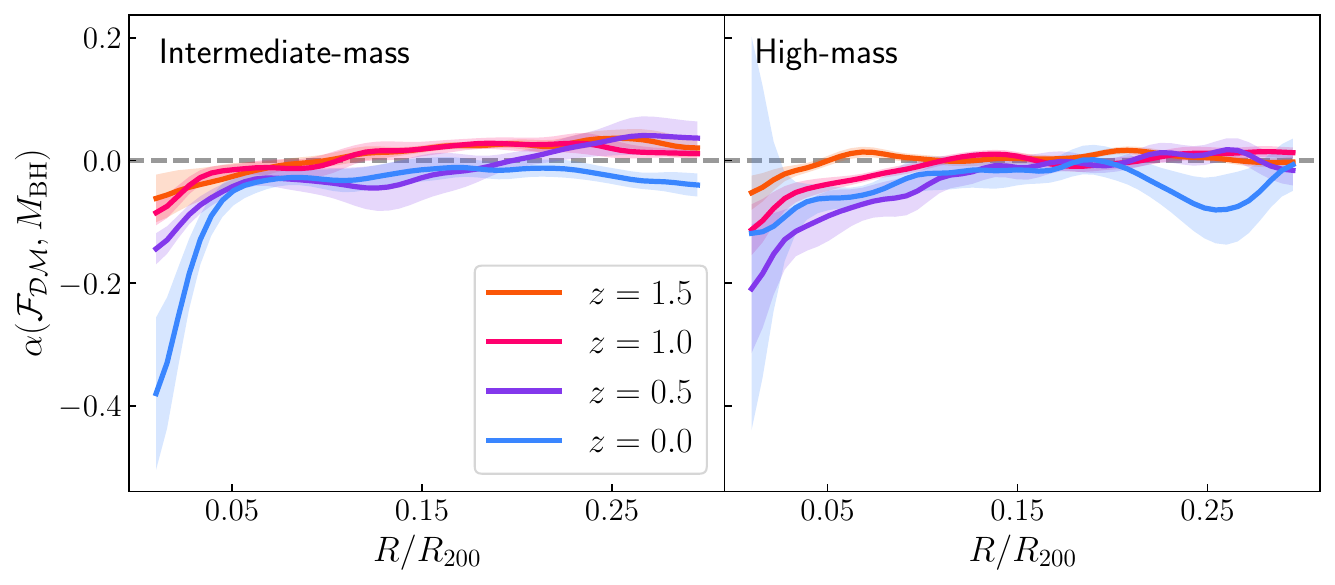}
     \caption{Variation of the slope, $\alpha$, of the $\mathcal{F_{DM}}-M_{\rm BH}$  relation at different fractions of $R_{\rm 200}$. 
    The relation is measured at $z=1.5$ (orange), $z=1.0$ (pink), $z = 0.5$ (purple), and $z=0.0$ (blue). 
    The shaded region represents the error in the slope calculated using 500 bootstrap runs. 
    The black dashed line has a null slope corresponding to a negligible AGN feedback effect on DM mass.
    The left and right panels highlight the intermediate and high mass bins.}
    \label{fig:slope_variation}
\end{figure*}

In this section, we have assessed the ``sphere of influence'' of the SMBH, by quantifying the spatial disturbance of AGN feedback on DM halos. 
Since low-mass galaxies show no difference in their DM suppression (\sec{dmfraction}), we shall only focus on the intermediate-mass and high-mass systems.

To quantify the spatial influence of the SMBH feedback, we introduce a parameter $\alpha(\mathcal{F_{DM}}, M_{\rm BH})$ defined as the slope of a linear fit between \fdm(R) and $M_{\rm BH}$.
The slope of this fit informs us about the interdependence between AGN feedback strength and its ability to create a DM suppression (via gravitational potential variations).
We would accordingly expect $\alpha(\mathcal{F_{DM}}, M_{\rm BH})$ to be negative at small radii and approach zero in the galaxy's outskirts.
Some examples of orthogonal linear fits at different redshifts and radii are presented in \Fig{fdm_mbh} in \app{alpha_def}.
Indeed at smaller radii, $\alpha(\mathcal{F_{DM}}, M_{\rm BH})$ (left-hand panel of \Fig{fdm_mbh}) is more negative than at larger galactocentric radii (right-hand panel of \Fig{fdm_mbh}).
As the transition in $\alpha(\mathcal{F_{DM}}, M_{\rm BH})$ is smooth as a function of $R/R_{\rm 200}$, we can also identify a turnover radius.

To avoid numerical resolution issues of the DM particles, the $\alpha(\mathcal{F_{DM}}, M_{\rm BH})$ values are calculated for $R/R_{\rm 200}>0.01$.
Furthermore, given the large softening lengths of SMBH particles, some of the results presented here may be attributed to numerical heating yielding artificial DM density suppression. 
We find that for intermediate-mass galaxies the median softening length for central SMBH at $z=0$ is $\sim$3\,kpc, while high-mass systems have a softening length of $\sim 10$\,kpc.
Therefore, the spheres of influence with size smaller than the typical softening lengths (\Fig{smbh_evol}) are likely caused by AGN feedback along with gravitational softening effects.
Those sizes are also related to the strength of the AGN feedback. 
In \app{epsilon_test}, we present \fdm profiles for NIHAO galaxy \texttt{g8.26e11} while varying the AGN feedback efficiency.
As expected, varying $\epsilon_{\rm f}$ changes the central SMBH mass and the softening lengths.

\Fig{slope_variation} shows $\alpha(\mathcal{F_{DM}}, M_{\rm BH})$ as a function of $R/R_{\rm 200}$ for the intermediate-mass (left-hand panel) and the high-mass (right-hand panel) systems at different redshifts (shown as different colours).
The value of $\alpha(\mathcal{F_{DM}}, M_{\rm BH})$ is negative for small radii, while approaching $\alpha(\mathcal{F_{DM}}, M_{\rm BH})\sim 0$ for larger radii.
This is true for all redshifts and both stellar mass bins.
The DM suppression is correlated with SMBH mass and connected to the overall gravitational potential.
For the intermediate-mass systems, $\alpha(\mathcal{F_{DM}}, M_{\rm BH})$ in the inner parts also grows over time.
By present day ($z=0$), \fdm and $M_{\rm BH}$ are strongly correlated; indeed, more massive SMBHs can create a larger DM suppression (\fdm$\sim -0.2$).

For the high-mass systems, similar trends are found albeit with smaller values for $\alpha(\mathcal{F_{DM}}, M_{\rm BH})$ at small radii. 
The smaller amplitude of $\alpha(\mathcal{F_{DM}}, M_{\rm BH})$ is related to the lack of SMBH mass evolution for the high-mass galaxies (see \Fig{smbh_evol}), coupled with the larger gravitational potential in massive galaxies counteracting the black hole feedback.
Interestingly, the strongest correlation ($\alpha(\mathcal{F_{DM}}, M_{\rm BH})$) is found at $z=0.5$, a feature also observed in \Fig{dm_ratio}.
With the lack of SMBH mass evolution and little gas content in the central parts, the AGN feedback is no longer effective and the DM halos in high-mass bins show signatures of halo contraction.

We have fit the $\alpha(\mathcal{F_{DM}}, M_{\rm BH})$ versus $R/R_{\rm 200}$ curves with a piece-wise linear function to calculate the sphere of influence, described as the turnover radius $R_{\rm t}$ of the fit, for black hole feedback on the DM.
These results are presented in \Table{turnover}.
Interestingly, for the intermediate-mass NIHAO galaxies, the sphere of influence is constant (within the errors) at $\sim 0.05\,R_{\rm 200}$ starting at $z=1.0$.
In physical units, the sphere of influence drops 
from $\sim$16\,kpc at $z=1.5$ to roughly $\sim$8\,kpc at present day. 
At present day, $R_{\rm t}=7.80\pm0.69\,{\rm kpc}$ is approximately 2 times the disk scale length of Milky Way-type discs \citep[][and references therein]{Rix2013}.
The sphere of influence encompasses approximately 80$\pm$23 per cent of the total stellar mass for intermediate mass galaxies.

\begin{table}
\begin{tabular}{@{}cccc@{}}
\toprule
Stellar mass bin                   & Redshift & $R_{\rm t}/R_{\rm 200}$ & $R_{\rm t}\,{[\rm kpc]}$ \\ 
(1)                                & (2)      & (3)                     & (4)                     \\ \midrule
Intermediate-mass                  & 1.5      & $0.08\pm0.05$           & 16.09$\pm$3.23           \\
                                   & 1.0      & $0.05\pm0.01$           & 10.95$\pm$1.02           \\
                                   & 0.5      & $0.04\pm0.01$           & 8.84$\pm$0.68           \\
                                   & 0.0      & $0.04\pm0.01$           & 7.80$\pm$0.69          \\ \midrule
High-mass                          & 1.5      & $0.05\pm0.01$           & 15.48$\pm$2.81           \\
                                   & 1.0      & $0.05\pm0.03$           & 16.66$\pm$2.44           \\
                                   & 0.5      & $0.05\pm0.04$           & 16.74$\pm$2.16           \\
                                   & 0.0      & $0.03\pm0.05$           & 9.07$\pm$1.23           \\ \bottomrule
\end{tabular}
\caption{Linear piece-wise fits for $\alpha(\mathcal{F_{DM}}, M_{\rm BH})$ versus $R/R_{\rm 200}$ for the NIHAO systems. 
Columns (1-2) give the stellar mass bin and redshift range, respectively, while columns (3-4) give the turnover radius in $R_{\rm 200}$ and physical units, respectively.}
\label{tab:turnover}
\end{table}

On the other hand, the high-mass NIHAO systems have an approximately constant sphere of influence from $z=1.5$ to $z=0.5$ (both in physical units and in terms of R$_{\rm 200}$), finally decreasing to $\sim 9$\,kpc at $z=0$.
It should be noted that scatter for $R_{\rm t}/R_{\rm 200}$ increases with redshift as well. 
Indeed, the AGN feedback and DM halo contraction due to baryons can balance out leading to a constant evolution of the sphere of influence.
Finally, at $z=0$ the size of the sphere of influence is approximately the same size as the median numerical softening for the central SMBH and should be interpreted with caution.

\section{Summary and Conclusions}
\label{sec:conclusion}

The coupling of SMBH (AGN) feedback and baryons in galaxies is a subject of continued enlightenment
\citep{Baldry2006, Croton2016, Lovell2018, Waterval2022}.
Here, we have investigated the response of the inner DM halo due to AGN feedback in simulated NIHAO galaxies.
The outflow and inflow of gas from the inner parts of galaxies due to feedback are expected to cause rapid fluctuations of the central gravitational potential leading to changes in DM particle orbits \citep{Governato2010, Pontzen2012}.
The signatures of these varying orbits are imprinted in the internal DM halo mass and density profiles \citep{Martizzi2013, Peirani2017, Maccio2020, Dekel2021}.
Using NIHAO zoom-in cosmological simulations with and without AGN feedback, we have quantified (i) the amount of DM supressed from the central parts of AGN hosting galaxies, and (ii) the spatial extent of the DM suppression (see \Eq{fdm}) due to AGN feedback.
Indeed, the presence of a central supermassive black hole for galaxies with $\log(M_{*}/M_{\odot})>10$ creates a DM suppression in the inner parts.
The DM suppression due to AGN feedback is a cumulative effect, being negligible at $z=1.5$ and growing to a value of \fdm$\sim -0.2$ ($\sim$40\,per\,cent) at present. 
At $z=0$, the DM suppression as a function of stellar mass is approximately constant (with \fdm$(R=0.02\,R_{\rm 200})\sim -0.2$), indicative of SMBH mass evolution in massive galaxies coupled with the gas availability for black hole accretion.
Indeed, for the high-mass galaxies the largest values of \fdm are found at $z=0.5$ after which, due to lack of SMBH mass evolution (therefore AGN feedback), the central parts of DM halos can contract again and regain their cuspy nature at $z=0$.
The results presented here for NIHAO are consistent with the Horizon-AGN and Illustris simulations which also find flattened DM density profiles at high redshift 
and suppression of the central DM mass due to AGN feedback for massive halos at all redshifts \citep{Peirani2017, Wang2019, Espinosa2022}.

The mass evolution of multiple galactic components (SMBH, cold gas, stellar mass, DM) was also characterized.
While intermediate-mass galaxies grow their SMBH mass monotonically until now, the growth in SMBH mass for high-mass systems saturates beyond $z\sim$1.5 and host a SMBH with median mass $\log(M_{\rm BH}/M_{\odot})\sim 8.5$ (see \Fig{smbh_evol}).
For baryons, the high-mass systems exhibit largest differences between the \agn and \noagn simulations. 
The amount of gas present in the central parts of high-mass \agn galaxies compared to \noagn is approximately two orders of magnitude lower.
With the fuel for SMBH accretion negligible, the absence of central black hole growth is expected.
This results in signatures of DM halo contraction for $z<0.5$ in the high-mass NIHAO systems. 
On the other hand, the adiabatic halo contraction due to the cooling baryons in \noagn simulations draws in more DM particles (as seen in the idealized simulations of \citealt{Martizzi2013}), making the baryonic-to-DM mass fraction sub-maximal in the central parts of these galaxies \citep{CourteauRix1999, Martinsson2013, Courteau2014, Courteau2015, Lovell2018}.

We have also constrained the SMBH sphere of influence, a spatial region within which the DM suppression and SMBH activities are correlated.
The slope, $\alpha(\mathcal{F_{DM}}, M_{\rm BH})$, was found to be steeper with time (from $z=1.5$ to now) for intermediate-mass galaxies.
This trend also holds for high-mass galaxies, though with slightly shallower slopes due to weaker black hole feedback at late times.
The most negative slope is found at $z=0.5$ for the high-mass system.
Using piece-wise linear regression, we calculated the radius within which black hole feedback is most effective at removing DM.
For both intermediate and high stellar mass bins, the sphere of influence remains constant (in units of virial radius) over time with an average value of ${\sim}0.05\,R_{\rm 200}$ or $\sim$8\,kpc\,h$^{-1}$, albeit at different time frames.
When measured in physical sizes, the sphere of influence decreases over the evolution of intermediate-mass galaxies as a result of the competition between the AGN feedback and total mass growth.
For high-mass systems, the sphere of influence remains constant until $z=0.5$; indeed at present day, the sphere of influence reaches a value lower than the resolution of the SMBH.
The spheres of influence presented, ranging from 3-5\,per cent of $R_{\rm 200}$, are sufficiently large to encompass a significant fraction 
(${\sim}80$ per~cent) of the total stellar content of simulated NIHAO galaxies. 
Our results provide evidence for the strong impact of AGN feedback on the evolution of DM halo (and baryonic) properties with time.
Indeed, NIHAO galaxies experience overall halo expansion as a result of the stellar feedback and dynamical friction \citep{Johansson2009, Cole2011, Dutton2016}.
With additional black hole feedback, the DM halo expansion for massive NIHAO galaxies is also amplified. 

\section*{Data Availability}
The data underlying this article will be shared upon request to the corresponding author(s).

\section*{Acknowledgements}

This material is based upon work supported by Tamkeen under the NYU Abu Dhabi Research Institute grant CASS.
SC is grateful to the Natural Sciences and Engineering Research Council of Canada, the Ontario Government, and Queen's University for generous support through various scholarships and grants.
The authors gratefully acknowledge the Gauss Centre for Supercomputing e.V. (www.gauss-centre.eu) for funding this project by providing computing time on the GCS Supercomputer SuperMUC at Leibniz Supercomputing Centre (www.lrz.de).
This research was carried out on the high performance computing resources at New York University Abu Dhabi. 
We greatly appreciate the contributions of all these computing allocations.
Vardha Bennert, Avishai Dekel, Arianna Di~Cintio, Jonathan Freundlich, Matteo Nori, and Romain Teyssier are thanked for insightful discussions and references.  
Vardha Bennert also kindly contributed observational literature for \Fig{mstar_mbh}.
We are also most grateful to an anonymous referee for a detailed review and suggesting valuable improvements.
The research was performed using the {\scriptsize PYNBODY} package \citep{Pontzen2013}, S{\scriptsize CI}P{\scriptsize Y} \citep{scipy}, and N{\scriptsize UM}P{\scriptsize Y} \citep{numpy} and used {\scriptsize MATPLOTLIB} \citep{matplotlib} for all graphical representation.


\bibliographystyle{mnras}
\bibliography{references} 



\appendix
\section{AGN Feedback Efficiency}\label{sec:epsilon_test}

\begin{figure*}
    \includegraphics[width=\linewidth]{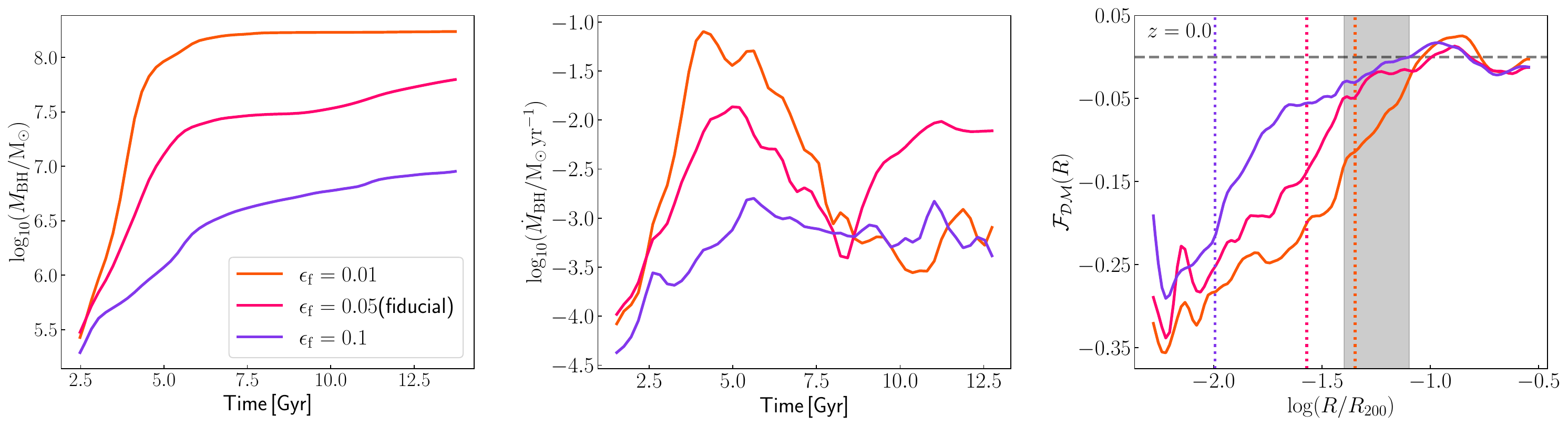}
    \caption{Left-hand panel: Evolution of the central black hole mass as a function of time for the \texttt{g8.26e11} simulation using different AGN feedback efficiency (shown using different line colours). 
    Centre panel: SMBH accretion rates as a function of time for the \texttt{g8.26e11} simulation using different AGN feedback efficiency (shown using different line colours).
    Right-hand panel: \fdm as a function of $R/R_{\rm 200}$ at $z=0$ for the \texttt{g8.26e11} simulation with different AGN feedback efficiency ($\epsilon_{\rm f}$). 
    The grey shaded region shows the distribution of the sphere of influence calculated for intermediate-mass simulated galaxies in \Table{turnover}. 
    The vertical dotted lines show the softening lengths for the central SMBH in each simulation (corresponding to the colour).}
    \label{fig:epsilon_f}
\end{figure*}

To test that the sphere of influence and DM suppression presented in this study are not due to resolution of the black hole particles in the simulation, we present the evolution of the central black hole mass (left-hand panel), black hole accretion rates (centre panel) and radial \fdm profile for the different AGN feedback efficiency in \Fig{epsilon_f}.
To carry out such resolution tests, we varied the AGN feedback efficiency ($\epsilon_{\rm f}$) by an order of magnitude centered around the fiducial value $(\epsilon_{\rm f} = 0.05)$ used for the NIHAO simulations.
A strong AGN feedback efficiency yields a smaller central SMBH mass.
This is expected as the strong feedback leads to strong outflow and smaller accretion rates (centre panel in \Fig{epsilon_f}) preventing black hole growth. 
The central SMBH mass changes approximately by an order of magnitude as a function of the AGN feedback efficiency within the range considered here.

The centre panel in \Fig{epsilon_f} shows the SMBH accretion rates for the three simulations as a function of time.
All three simulations present the peak $\log(\dot{M}_{\rm BH}/{\rm M_{\odot}\, yr^{-1}})$ at similar times ($t\sim 5\,{\rm Gyr}$).
The SMBH with the lowest AGN feedback efficiency results in the largest accretion rates and consequently the largest black hole mass.
The right-hand panel of \Fig{epsilon_f} shows the \fdm profiles for \texttt{g8.26e11} with varying AGN feedback efficiency 
(with different colours).
The vertical lines in the right-hand panel of \Fig{epsilon_f} represent the softening lengths for the central SMBH particle which scale with the SMBH mass.
The central SMBH with the largest accretion rates (albeit with the lowest $\epsilon_{\rm f}$) is sufficiently energetic to produce the largest \fdm.
For the Bondi-Hoyle-Lyttleton parametrization, the instantaneous energy output of the AGN feedback is linearly proportional to the AGN feedback efficiency and SMBH accretion rates (or $\propto M_{\rm BH}^2$).
Therefore, the instantaneous energy output (and therefore, the DM suppression) is more sensitive to the central SMBH mass than to the AGN accretion efficiency.
The lowest-mass SMBHs (corresponding to $\epsilon_{\rm f}=0.1$) create the smallest DM suppression while the highest-mass SMBHs (due to large$\dot{M}_{\rm BH}$) give rise to large DM suppression.
Indeed, all softening lengths in \Fig{epsilon_f} are smaller than the sphere of influence in intermediate-mass NIHAO simulations (shown in \Table{turnover} and the gray shaded region in \Fig{epsilon_f}), thus minimizing the impact of numerical resolution on our results.
The highest mass SMBH does have a softening length on the lower-end of the sphere of influence indicating the partial role that resolution plays in setting DM suppression at high masses.

\section{SMBH and Dark Matter Deficit}\label{sec:alpha_def}

\begin{figure*}
    \centering
    \includegraphics[width=\linewidth]{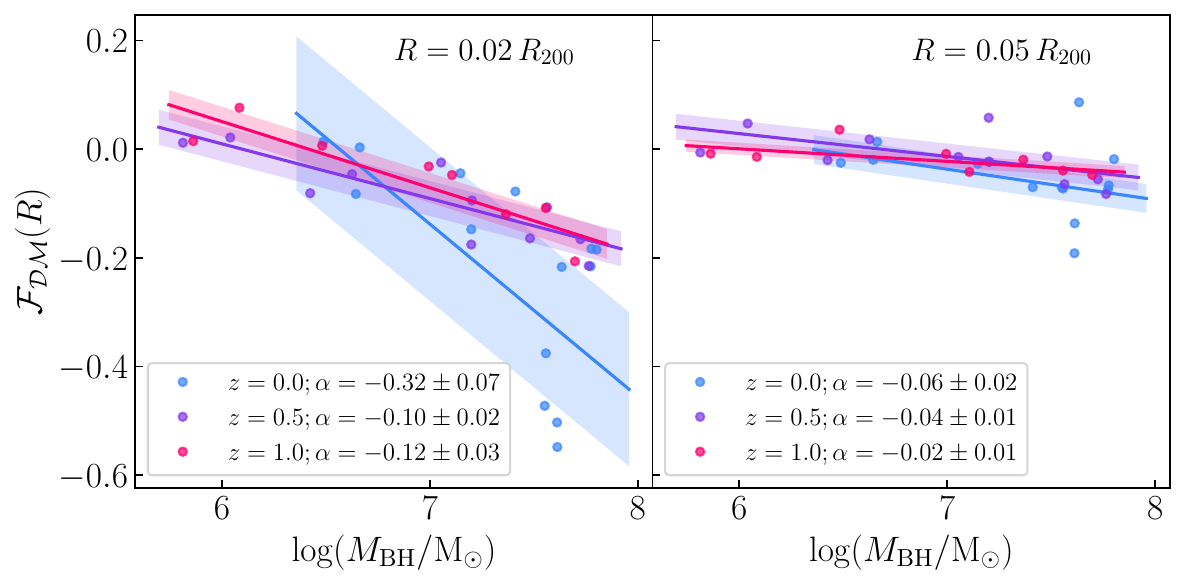}
    \caption{\fdm versus $\log(M_{\rm BH}/{\rm M_{\rm \odot}})$ for intermediate-mass NIHAO galaxies at three different redshifts. 
    The quantity \fdm is calculated at 0.02\,R$_{\rm 200}$ (left-hand panel) and 0.05\,R$_{\rm 200}$ (right-hand panel). 
    The data points show simulated NIHAO galaxies, while the solid line and shaded regions represent the best linear fits (slope and scatter). 
    The slopes ($\alpha(\mathcal{F_{DM}}, M_{\rm BH})$) and their error for each linear fit are given in each legend. 
    }
    \label{fig:fdm_mbh}
\end{figure*}

This Appendix shows examples of the inferred parameter $\alpha(\mathcal{F_{DM}}, M_{\rm BH})$, which is used to calculate the sphere of influence of AGN feedback on the DM halo of galaxies.
Recall that $\alpha$ is the slope of the $\mathcal{F_{DM}}-M_{\rm BH}$ relation (see \sec{bh_influence}).  
\Fig{fdm_mbh} shows \fdm measured at 0.02\,R$_{\rm 200}$ (left-hand panel) and 0.05\,R$_{\rm 200}$ (right-hand panel) versus SMBH mass for NIHAO galaxies with $10.0\leq\log(M_*/{\rm M_{\odot}})<10.5$. 
An orthogonal linear best fit is performed at each redshift to calculate $\alpha(\mathcal{F_{DM}}, M_{\rm BH})$ which relates the interdependence between the dark matter deficit and SMBH mass.
As expected, smaller values of $\alpha(\mathcal{F_{DM}}, M_{\rm BH})$ and minimal variations with redshift are found at larger galactocentric radii ($R = 0.05\,R_{\rm 200}$). 
The larger negative values of $\alpha(\mathcal{F_{DM}}, M_{\rm BH})$ at smaller galactocentric radii highlight the local impact of AGN feedback on dark matter halos in galaxies.
This effect is most noticeable for local galaxies ($z\sim0$) at smaller radii.

\bsp	
\label{lastpage}
\end{document}